\documentclass[a4paper,11pt]{article}

\usepackage{amsmath}
\usepackage[dvips]{graphicx}

\def\D{\nabla}

\def\U{\mathcal{U}}

\def\ch{\rm{ch}}
\def\sh{\rm{sh}}

\renewcommand{\theequation}{\arabic {section}.\arabic{equation}}
\makeatletter
\@addtoreset{equation}{section}
\makeatother

\begin{document}

\begin{flushright}
February, 2025
\end{flushright}

\vspace{0.0cm}

\begin{center}

{\Large \bf Non(anti)Commutative Superspace,  \\
\vspace{0.3cm}
Baker-Campbell-Hausdorff Closed Forms, and \\
\vspace{0.5cm}
Dirac-K\"ahler Twisted Supersymmetry}

\vspace{1cm}

{ Kazuhiro Nagata}\footnote{khnagata@yahoo.co.jp}
\\


\end{center}


\begin{abstract}
Starting from an elementary calculation of super Lie group elements associating with 
non(anti)-commutative Grassmann parameters, we derive several closed expressions of Baker-Campbell-Hausdorff (BCH)  formula which represent multiplication properties of super Lie group elements in the corresponding superspace.
We then show that parametrization of superspace in general may become infinite dimensional due to the presence of non(anti)commutativity.
We show that a Dirac-K\"ahler Twisted SUSY Algebra (also referred to as Marcus B-type Twisted SUSY Algebra or Geometric Langlands Twisted SUSY Algebra) with 
a certain type of deformation, which we call an exponential deformation,
 may circumvent this problem.
We also provide, in terms of gauge covariantization of the SUSY algebra,
 a geometric understanding of the exponential deformation,
and see that the framework constructed in this paper may serve as 
a non(anti)commutative superspace framework providing
a recently proposed gauge covariant link formulation of twisted super Yang-Mills
on a lattice.
\end{abstract}


\section{Introduction}

\indent

Since more than a few decades ago, relations between gauge theory and non-commutative geometry have been attracting a lot of attentions 
\cite{Connes, Connes_Douglas_Schwarz, Seiberg_Witten, Minwalla, Schwarz, Li_Wu},
and 
many related studies have been carried out
mainly motivated by string theory, or
in pursuing more granular structure of spacetime
(for a recent introduction to this subject, see, for example,  
\cite{Vitale} and references therein).
At the early stage of these studies, non-commutative properties of the geometry were examined mainly
in the bosonic sector of the  coordinates, $x_{\mu}$, which obey,
\begin{eqnarray}
[x_{\mu},x_{\nu}] &=& i\Theta_{\mu\nu}, 
\end{eqnarray}
where $\Theta_{\mu\nu}$, which is anti-symmetric w.r.t. the vector indices $\mu$ and $\nu$, represents the degree of non-commutativity.
The scope was then naturally extended to non-commutativities, or more precisely  non(anti)commutativities including the one between fermionic coordinates, $\theta^{\alpha}$ \cite{Klemm},
\begin{eqnarray}
\{\theta^{\alpha},\theta^{\beta} \} &=& C^{\alpha\beta}, \label{NACR1}
\end{eqnarray}
where
$C^{\alpha\beta}$, which is symmetric w.r.t. the spinor indices $\alpha$ and $\beta$,
  represents the degree of non(anti)commutativity.
It is then understood that the above type of non(anti)commutativity between fermionic coordinates naturally arises as an effect of  graviphoton background \cite{deBoera, Ooguri_Vafa, Seiberg, Berkovits_Seiberg, Ivanov}. 
Due to an existence of non-commutativity between the coordinates, Poincar\'e symmetry of the spacetime in the ordinary sense may be broken. Therefore, investigating how to conceptually recover the  Poincar\'e symmetry in the non-commutative spacetime
also attracted attentions. One of the ideas was given by  the deformation of  Hopf algebras \cite{Drinfeld} and provided  deformations of Poincar\'e algebra  by  introducing so-called twist elements \cite{Chaichian, Oeckl, Kobayashi_Sasaki}\footnote{
Note that the term ``twist'' in the above context represents a deformation of Hopf algebras,
whereas in most of the parts of this paper, the 
``twist'' represents a topological twisting \cite{Witten1} of supersymmetry (SUSY) algebra.}.

As mentioned above, historically, non(anti)commutativity between fermionic coordinates
has been examined mainly in the chiral or anti-chiral sector,
not in a vector sector,
In other words. the non(anti)commutativity of the  fermionic coordinates
has been introduced as $C^{\alpha\beta}$ as in (\ref{NACR1})  or 
$C^{\dot{\alpha}\dot{\beta}}$ which may respectively correspond to $(\bf{1}, \bf{0})$ 
or $(\bf{0},\bf{1})$ representation of the Lorentz group,
not a representation $(\bf{\frac{1}{2}},\bf{\frac{1}{2}})$
where the momentum operator resides via the SUSY algebra, 
$\{Q_{\alpha},\overline{Q}_{\dot{\beta}}\} = P_{\alpha\dot{\beta}}$.

In this paper, 
we mainly examine a non(anti)commutativity in the vector sector.
Namely, 
in terms of a generic notation, we deal with a situation
where a non(anti)commutativity
\begin{eqnarray}
\{\theta_{A},\theta_{B} \} &=& a_{AB} 
\end{eqnarray}
is associating with the corresponding SUSY algebra,
\begin{eqnarray}
\{Q_{A},Q_{B} \} &=& P_{AB}.
\end{eqnarray}
Although the motivation of this study initially stemed from the author's 
rather primitive interest
of relations between SUSY and non-locality encompassing lattice SUSY,
the result of  the study turns to
include several non-trivial and practical applications including
Baker-Campbell-Hausdorff (BCH) exact formula for a certain type of commutation 
relations, and a non(anti)commutative
superspace framework providing a realization of exact SUSY on a lattice.

The content of this paper is as follows.
In section 2, introducing non(anti)commutative Grassmann parameters,
and investigating multiplication properties of super Lie group elements, we see
that the corresponding algebra may be infinite dimensional due to the presence of non(anti)commutativity.
In section 3, we discuss how to circumvent the emergence of the infinite dimensionality in the suprespace and then propose a certain type of  deformation of
SUSY algebra, which we call in this papar as an exponential deformation.
We then see that in order to consistently perform the exponential deformation,
a certain type of conditions, which we have been calling as lattice Leibniz rule
conditions, should be satisfied.

In section 4,
as an example of SUSY algebra which satisfy the lattice Leibniz rule
condition, we explain Dirac-K\"ahler SUSY Algebra (also referred to as Marcus B-type or Geometric Langlands Twisted SUSY Algebra) in various dimensions and their exponential deformation..
We then introduce a certain type of Ansatz on the non(anti)commutativity 
between the Grassmann coordinates, with which the non-linearity 
appeared in the multiplication properties among the super Lie group element
turns to be governed by a set of ratio parameters. One of these parameters is a ratio between
the above-mentioned $a_{AB}$
and a lattice constant $n_{AB}$.

In section 5, we pick up several particular cases of 
non(anti)commutative problem settings and see that  
they
give rise to
 feasible supertranslation properties in the superspace.
In section 6, in terms of gauge covariantization of the SUSY algebra,
we provide a geometric understanding of exponential deformation,
and see that the framework constructed in this paper may serve as 
a non(anti)commutative superspace framework providing
a recently proposed gauge covariant link formulation of twisted super Yang-Mills
in various dimenisons \cite{DKKN2, DKKN3, DKNS}.
We provide several comments and discussions for future perspectives in section 7.
An explicit derivation of Baker-Campbell-Hausdorff Closed Forms
by means of non(anti)commutative Grassmann parameters and supercharges are shown in the Appendix.

\section{SUSY Algebra, BCH formula, and Inifinite-dimensionality}

\indent

In this section, after introducing non(anti)commutative Grassmann parameters, we investigate multiplication properties of super Lie group elements. The calculations are based on Baker-Campbell-Hassdorff closed forms derived in the Appendix.
We then explicitly show the 
parametrization of superspace in general may become inﬁnite dimensional due to the presence of non(anti)commutativity,
which may be understood as an obstacle to preserving supersymmetry in the present non(anti)commutative settings.


Let us begin with the following SUSY algebra and Grassmann parameters with non(anti)-commutativity,
\begin{eqnarray}
\{Q_{A},Q_{B}\} &=& P_{AB}, \ \ \ Q_{A}^2 \ =\ Q_{B}^2 \ =\ 0, \label{SUSY_alg} \\ 
\{\theta_{A},\theta_{B}\} &=& a_{AB}, \ \ \ \ \theta_{A}^2 \ =\ \theta_{B}^2 \ =\ 0. \label{NAC}
\end{eqnarray}
Here, the symbol $\{\cdot, \cdot\}$ stands for an anti-commutator defined by
$\{a, b\} \equiv ab + ba$, and  we use a generic notation where $Q_{A}$ and $Q_{B}$ denote supercharges while $\theta_{A}$ and $\theta_{B}$ denote Grassmann parameters whose degree of non(anti)commutativity is represented by $a_{AB}$. The Grassmann parameters $\theta_{A}$ and $\theta_{B}$ may also be regarded as non(anti)commutative Grassmannn coordinates which span the target non(anti)commutative superspace. $P_{AB}$ denotes a momentum operator. The subscripts $A$ and $B$ may represent any indices depending on how we describe spacetime symmetries of the supercharges and extended supersymmetries. For example, the subscripts  $A$ and $B$ may be dotted and undotted spinor indices in a case of ordinary (untwisted) supersymmetry, while they may be scalar, vector, or tensor indices in a case of topologically twisted supersymmetry.
Here, we assume that $a_{AB}$ may be any c-number. 
Throughout this article, we treat the supercharges $Q_{A}$ and $Q_{B}$ as abstract group generators in the sense of \cite{Gates_Grisaru_Rocek_Siegel} by assuming anti-commutativities between $Q_{A}$'s and $\theta_{A}$'s, namely,
\begin{eqnarray}
\{\theta_{A}, Q_{A}\} &=& \{\theta_{B}, Q_{B}\} \ =\ \{\theta_{A}, Q_{B}\} \ =\ \{\theta_{B}, Q_{A}\} \ =\ 0.
\ \ (A, B:\rm{no\ sum})
\end{eqnarray}

\subsection{Multiplication properties between one-parameter group elements}

\indent

In this subsection, as a first example, we examine multiplication properties between one-parameter group elements belonging to a super Lie group whose corresponding algebra consisting of $Q_{A}, Q_{B}, P_{AB}$. Strictly speaking, as mentioned above, the parameters $\theta_{A}'s$ are not ordinary parameters, but they should be treated as elements of a non(anti)commutative ring. 
In this regard, the group elements treated in this subsection should be expressed as
``would-be one-parameter group elements'' which reduce to ordinary super Lie
group elements in the anti-commutative limit $a_{AB}\rightarrow 0$. 
Although, in general, these elements may be expressed in terms of the language of tensor  algebra as
$e^{C\otimes D+\cdots}$ with a graded commutatitity
$C\otimes D = (-1)^{|C||D|}D\otimes C$, where the degree $|C|$ takes $0$ (or $1$) for  bosonic  $C$ (or fermionic), and the same applying for $D$,
 we omit the symbol $\otimes$ in the following.

In constructing the one-parameter group elements, we 
examine a product of a SUSY generator $Q_{A}$ and a corresponding coordinate $\theta_{A}$ spanning a target superspace. Note here that the meaning of the coordinates of the target superspace is a bit obscure at this stage, since we do not have any a priori knowledge of entire set of generators consistent with multiplication rules of super the Lie group elements.
Nevertheless, as a first example, if we define,
\begin{eqnarray}
X \equiv \theta_{A}Q_{A}, \ \ \ \ Y \equiv \theta_{B}Q_{B}, \ \ \ \ (A, B:\rm{no\ sum}) \label{XY}
\end{eqnarray}
we have, as a commutation relation of these objects,
\begin{eqnarray}
[X,Y] &=& \theta_{A}Q_{A} \theta_{B}Q_{B} - \theta_{B}Q_{B}\theta_{A}Q_{A}\\[5pt]
&=& - \theta_{A}\theta_{B}Q_{A}Q_{B} + (-\theta_{A}\theta_{B} + a_{AB})Q_{B}Q_{A} \\[5pt]
&=& - \theta_{A}\theta_{B} P_{AB} + a_{AB}Q_{B}Q_{A}. \label{XY1}
\end{eqnarray}
Here, the second term in (\ref{XY1}) stems from the non(anti)commutativity
between $\theta_{A}$ and $\theta_{B}$ and includes an element $Q_{B}Q_{A}$
which resides outside of the 
starting set of  algebraic elements $Q_{A}$, $Q_{B}$, and $P_{AB}$.
Before going further, it is worthwhile to mention that the first and second terms in (\ref{XY1}) are complementary each other in a sense that $\theta_{A}\theta_{B}$ can be regarded as an idempotent object with an ``eigenvalue'' $a_{AB}$, while $Q_{A}Q_{B}$ can be regarded as an idempotent object with an ``eigenvalue(eigen-operator)'' $P_{AB}$,
\begin{eqnarray}
(\theta_{A}\theta_{B})^2 &=& a_{AB}\theta_{A}\theta_{B}, \\[5pt]
(Q_{B}Q_{A})^2 &=& P_{AB}Q_{B}Q_{A}.
\end{eqnarray}
More particularly, if we consider an exponentiation of each of these object, we have,
\begin{eqnarray}
e^{\theta_A \theta_B P_{AB}} 
&=& 1 +  \theta_A \theta_B P_{AB} +  \frac{1}{2!} (\theta_A \theta_B)^2 P_{AB}^2 
+  \frac{1}{3!} (\theta_A \theta_B)^3 P_{AB}^3 +  ... \\[5pt]
&=& 1 +  \frac{\theta_A \theta_B}{a_{AB}} \Bigl( e^{a_{AB}P_{AB}} -1 \Bigr) \label{finite0}
\end{eqnarray}
which can be expressed as
\begin{eqnarray}
a_{AB}\Bigl(e^{\theta_A \theta_B P_{AB}}-1\Bigr) 
&=& \theta_{A}\theta_{B} \Bigl( e^{a_{AB}P_{AB}} -1 \Bigr), \label{aexp}
\end{eqnarray}
and, in a similar manner,
\begin{eqnarray}
P_{AB}\Bigl(e^{a_{AB}Q_{B}Q_{A}}-1\Bigr) 
&=& Q_{B}Q_{A} \Bigl( e^{a_{AB}P_{AB}} -1 \Bigr). \label{Pexp}
\end{eqnarray}
The appearance of the finite shift operators $e^{a_{AB}P_{AB}}$ in the r.h.s. of (\ref{aexp}) and (\ref{Pexp}) lead us to a plausible observation that the non(anti)commutativity $\{\theta_{A},\theta_{B}\}=a_{AB}$ may intrinsically be related to a
discretization of spacetime with a lattice constant of $O(a_{AB})$.
We will come back to this point later on.

Returning to the calculation of commutation relations of $X$ and $Y$, if  we take further commutation relations with $X$ and $Y$ respectively, we have,
\begin{eqnarray}
[X,[X,Y]] &=& [\theta_{A}Q_{A}, - \theta_{A}\theta_{B} P_{AB} + a_{AB}Q_{B}Q_{A}] \\[5pt]
&=& a_{AB}P_{AB}\theta_{A}Q_{A} + a_{AB}P_{AB}\theta_{A}Q_{A}\\[5pt]
&=& 2 a_{AB}P_{AB} X, \ \ \ \rm{and} \label{XXY1} \\[5pt]
[Y,[Y,X]] &=& 2 a_{AB}P_{AB} Y. \label{YYX1}
\end{eqnarray}
These commutation relations  (\ref{XXY1}) and (\ref{YYX1}) induce a multiplication property between two super Lie group elements
$e^{X}$ and $e^{Y}$ via the Baker-Campbell-Hausdorff (BCH) formula as follows
\begin{eqnarray}
e^{X} e^{Y}  &=&
\exp \Bigl[ X+Y + \frac{1}{2}[X,Y] + \frac{1}{12} [X,[X,Y]]  + \frac{1}{12} [Y,[Y,X]]  \nonumber \\ 
&&+ \frac{1}{24}[Y,[X,[X,Y]]]] + ...\Bigr] \\[5pt]
&=& \exp \Bigl[ \bigl(1+\frac{1}{6} a_{AB}P_{AB} + ... \bigr)X+ \bigl(1+\frac{1}{6} a_{AB}P_{AB} + ... \bigr)Y \nonumber \\[5pt] 
&&+ \bigl(1+\frac{1}{6} a_{AB}P_{AB} + ... \bigr)\frac{1}{2} [X,Y] \Bigr],
\end{eqnarray}
which strongly implies that this multiplication rule may be expressed as a closed form
\begin{eqnarray}
e^{X} e^{Y}   &=& \exp \Bigl[ f(a_{AB}P_{AB}) \big(X+Y+\frac{1}{2}[X,Y]\bigr) \Bigr]. \label{XY_BCH1}
\end{eqnarray}
Actually, one can see that the above function $f(a_{AB}P_{AB})$ turns out to be expressed as 
\begin{eqnarray}
f(a_{AB}P_{AB}) &=& \frac{\log \Bigl(1-\frac{a_{AB}P_{AB}}{2}+\sqrt{(1-\frac{a_{AB}P_{AB}}{2})^2-1}\Bigr)}
{\sqrt{(1-\frac{a_{AB}P_{AB}}{2})^2 -1}},  \label{XY_BCH2}
\end{eqnarray}
which is a special case of more general closed expressions of BCH formula derived in the Appendix.
Notice here that, as a function of the momentum operator $P_{AB}$, $f(a_{AB}P_{AB})$ takes a value of 1, only when $a_{AB}=0$.  
Together with the definitions of $X$ and $Y$ in Eq.(\ref{XY}), the relation (\ref{XY_BCH1}) tells us that, in a case of $a_{AB}\neq 0$, the supersymmetry generators $Q_{A}$ and $Q_{B}$ can not easily stand alone but tightly be connected with the momentum operator $P_{AB}$ 
whenever carrying out a super Lie group multiplication.

One may wonder, at this stage, that the contribution of $f(a_{AB}P_{AB})$ may be absorbed into re-definitions of $X$ and $Y$ (in other words, into re-definitions of $Q_{A}$ and $Q_{B}$). 
In order to carefully check such a possibility, it is worthwhile to take a look at a bit more general closed expression of BCH formula which can be obtained by replacing $X\rightarrow \sigma X$, $Y\rightarrow \tau Y$, $a_{AB}\rightarrow \sigma\tau a_{AB}$ in (\ref{XY_BCH1}), or via an explicit derivation as shown in the Appendix,
\begin{eqnarray}
e^{\sigma X} e^{\tau Y}   &=& \exp \Bigl[ f(\sigma\tau a_{AB}P_{AB}) \big(\sigma X+\tau Y+\frac{1}{2}\sigma\tau [X,Y]\bigr) \Bigr], \label{XY_BCH3}
\end{eqnarray}
where $\sigma$ and $\tau$ represent any commuting functions which may depend on $P_{AB}$, while $f(\sigma\tau a_{AB}P_{AB})$ is defined by Eq. (\ref{XY_BCH2}) by replacing $a_{AB}P_{AB}$ by
$\sigma\tau a_{AB}P_{AB}$ on the both sides.
Viewing the above relation (\ref{XY_BCH3}), one sees that the factor $f$ cannot be absorbed into this type of redefinitions,  since $\sigma X$ and $\tau Y$ appear on the both sides of the relation (\ref{XY_BCH3}),
and $f(\sigma\tau a_{AB}P_{AB})$, as a function of $P_{AB}$, 
never takes a value of 1 as long as $a_{AB}\neq 0$.

Another important aspect concerning the algebra of $X$ and $Y$
is
that the relations (\ref{XXY1}) and (\ref{YYX1}) induce
the following recursive relations,
\begin{eqnarray}
[X,Y] &=& 
- \theta_{A}\theta_{B} P_{AB} + a_{AB}Q_{B}Q_{A}
\ \equiv\  Z \\[5pt]
[X,[X,Y]] &=& [X,Z] \ =\ +2a_{AB}P_{AB} X \\[5pt] 
[Y,[X,Y]] &=& [Y,Z] \ =\ -2a_{AB}P_{AB} Y \\[5pt] 
[Y,[X,[X,Y]]] &=& +2a_{AB}P_{AB} [Y,X] \ = \ - 2a_{AB}P_{AB}Z \\[5pt]
[X,[Y,[X,Y]]] &=& -2a_{AB}P_{AB} [X,Y] \ = \ - 2a_{AB}P_{AB}Z \\[5pt]
[X,[Y,[X,[X,Y]]]] &=& -2a_{AB}P_{AB} [X,Z] \ = \ - (2a_{AB}P_{AB})^2 X \\[5pt]
[Y,[X,[Y,[X,Y]]]] &=& -2a_{AB}P_{AB} [Y,Z] \ = \ + (2a_{AB}P_{AB})^2 Y ,\ \ \  etc...
\end{eqnarray}
Therefore, if we define
\begin{eqnarray}
X_{n} &\equiv& (2a_{AB}P_{AB})^{n} X,\ \ \ \ \ n=0,1,2,...\\[5pt]
Y_{n} &\equiv& (2a_{AB}P_{AB})^{n} Y,\ \ \ \ \ n=0,1,2,...\\[5pt]
Z_{n} &\equiv& (2a_{AB}P_{AB})^{n} Z,\ \ \ \ \ n=0,1,2,...,
\end{eqnarray}
these generators form the following  infinite dimensional Lie Algebra,
\begin{eqnarray}
[X_{n},Y_{m}]&=&Z_{n+m}\\[5pt]
[X_{n},Z_{m}]&=&+X_{n+m+1}\\[5pt]
[Y_{n},Z_{m}]&=&-Y_{n+m+1},
\end{eqnarray}
from which we can read out that the entire superspace which accommodates the SUSY algebra (\ref{SUSY_alg}) attached with the non(anti)commutative Grassmann parameters (\ref{NAC})
may be spanned by an infinite number of coordinates, denoted as, for example,
\begin{eqnarray}
\theta_{A}^{(n)}, \ \ \ \theta_{B}^{(n)}, \ \ \ x^{(n+1)},\ \ \ (n=0, 1, 2, ...).
\end{eqnarray}
Each of these coordinates corresponds to each of
the infinite number of generators embedded in the above $X_{n}$, $Y_{n}$, and $Z_{n}$, as follows,
\begin{eqnarray}
(P_{AB})^n Q_{A},\ \ \ (P_{AB})^n Q_{B}, \ \ \  (P_{AB})^{n+1},\ \ \ (n=0, 1, 2, ...).
\end{eqnarray}
The above infinite dimensionality of the superspace 
implies that the above type of non(anti)commutativity,
 $\{\theta_{A},\theta_{B}\}=a_{AB}\neq 0$,
may in general break the corresponding SUSY in the ordinary sense.

\subsection{Multiplication properties between two-parameter group elements}

\indent

Let us continue to survey the multiplication properties of super Lie group elements. On the  superspace, in order to find out how the Grassmann 
coordinates 
vary under an action of super Lie group element, we need to introduce another set of Grassmann parameters 
other than 
the Grassmann coordinates $\theta_{A}$ and $\theta_{B}$.
Thus, as a second example of the super Lie group multiplication,
let us define,
\begin{eqnarray}
X \equiv \xi_{A}Q_{A} + \xi_{B}Q_{B}, \ \ \ \ Y \equiv \theta_{A}Q_{A} + \theta_{B}Q_{B},
\ \ \ \ (A,B:\rm{no\ sum}) \label{XY2}
\end{eqnarray}
where we introduced $\xi_{A}$ and $\xi_{B}$ as another set of non(anti)commutative Grassmann parameters which satisfy,
\begin{eqnarray}
\{\xi_{A},\xi_{B}\} &=& c_{AB}, \ \ \ \ \xi_{A}^2 \ =\ \xi_{B}^2 \ =\ 0,
\end{eqnarray}
where, in order to make the discussion as general as possible, we represent a degree of non(anti)-commutativity between $\xi_{A}$ and $\xi_{B}$ as $c_{AB}$ which may or may not be equal to 
$a_{AB}$, depending on a particular problem setting. 

Although anticommutativity between $\xi_{A}$'s and $\theta_{A}$'s may also depend on particular problem settings, here we impose,
\begin{eqnarray}
\{\xi_{A},\theta_{B}\} = \{\xi_{B},\theta_{A}\} =  b_{AB}, \ \ \ \ \{\xi_{A},\theta_{A}\} = \{\xi_{B},\theta_{B}\} = 0 \ \ \ \ (A, B:\rm{no\ sum}),
\end{eqnarray}
where a degree of non(anti)commutativity between $\xi_{A}$ and $\theta_{B}$ is represented by $b_{AB}$ which may or may not be equal to $a_{AB}$ or $c_{AB}$.
In the following, we will keep $a_{AB}$, $b_{AB}$ and $c_{AB}$ independent each other, unless otherwise mentioned.
In addition, as in the case of $\theta_{A}$'s, we assume,
\begin{eqnarray}
\{\xi_{A}, Q_{A}\} &=& \{\xi_{B}, Q_{B}\} \ =\ \{\xi_{A}, Q_{B}\} \ =\ \{\xi_{B}, Q_{A}\} \ =\ 0.
\ \ (A, B:\rm{no\ sum}).
\end{eqnarray}

Based on the above settings, as for a commutation relation of $X$ and $Y$ defined in Eq. (\ref{XY2}), we have,
\begin{eqnarray}
[X,Y] &=& [\xi_{A}Q_{A} + \xi_{B}Q_{B}, \theta_{A}Q_{A} + \theta_{B}Q_{B}] \\[5pt]
&=& [\xi_{A}Q_{A}, \theta_{A}Q_{A}] + [\xi_{A}Q_{A}, \theta_{B}Q_{B}] + [\xi_{B}Q_{B}, \theta_{A}Q_{A}] 
+ [\xi_{B}Q_{B}, \theta_{B}Q_{B}], 
\end{eqnarray}
here, the first and fourth terms are vanishing because of $\{\theta_{A},\xi_{A}\}=\{\theta_{B},\xi_{B}\} = Q_{A}^{2} = Q_{B}^{2} =0$, while the remaining terms give, 
\begin{eqnarray}
 [\xi_{A}Q_{A}, \theta_{B}Q_{B}] 
&=&  \xi_{A}Q_{A} \theta_{B}Q_{B} - \theta_{B}Q_{B}\xi_{A}Q_{A} \\[5pt]
&=& - \xi_{A}\theta_{B}Q_{A}Q_{B} + (-\xi_{A}\theta_{B} + b_{AB})Q_{B}Q_{A} \\[5pt]
&=& - \xi_{A}\theta_{B} P_{AB} + b_{AB}Q_{B}Q_{A}, \ \ \rm{and} \\[5pt]
 [\xi_{B}Q_{B}, \theta_{A}Q_{A}]
 &=& - \xi_{B}\theta_{A} P_{AB} + b_{AB}Q_{A}Q_{B},
\end{eqnarray}
thus, we have,
\begin{eqnarray}
[X,Y] &=& -( \xi_{A}\theta_{B} + \xi_{B}\theta_{A} -b_{AB})P_{AB}.
\end{eqnarray}
If  we take further commutation relations with $X$ and $Y$, respectively, after simple calculations, we obtain,
\begin{eqnarray}
[X,[X,Y]] 
&=& - c_{AB}P_{AB} Y + b_{AB}P_{AB}X, \ \ \rm{and} \label{XXY2} \\[5pt]
[Y,[Y,X]] 
&=& - a_{AB}P_{AB} X + b_{AB}P_{AB}Y. \label{YYX2} 
\end{eqnarray}
Note here that the relations (\ref{XXY2}) and (\ref{YYX2}) may be expressed as a general forms,
\begin{eqnarray}
[X,[X,Y]] &=& \gamma Y + \beta X, \ \ \rm{and} \label{XXY3} \\[5pt]
[Y,[Y,X]] &=& \alpha X + \beta Y, \label{YYX3} 
\end{eqnarray}
where
\begin{eqnarray}
\alpha &\equiv& - a_{AB}P_{AB}, \ \ \ \
\beta \ \equiv\ + b_{AB}P_{AB}, \ \ \ \ 
\gamma \ \equiv\ - c_{AB}P_{AB}.
\end{eqnarray}
Note also that the relations (\ref{XXY1}) and (\ref{YYX1})
which we obtain in the case of one-parameter group elements 
 can be regarded as 
a special case of (\ref{XXY3}) and (\ref{YYX3}) with $\alpha = \gamma = 0$ and $\beta = 2a_{AB}P_{AB}$.

The commutation relations (\ref{XXY2}) and (\ref{YYX2}) induce the following multiplication property between two super Lie group elements
$e^{X}$ and $e^{Y}$ by using the closed expression of the BCH formula derived in the Appendix,

\begin{eqnarray}
e^{X} e^{Y}   &=& \exp \Biggl[ F(\alpha, \beta, \gamma) \Big(G(\alpha) X 
+G(\gamma) Y +\frac{1}{2}[X,Y]\Bigr) \Biggr],  \nonumber 
\label{BCH_closed1}
\\
\end{eqnarray}
where $F(\alpha, \beta, \gamma)$, $G(\alpha)$, and $G(\gamma)$ are given by,

\begin{eqnarray}
&&F(\alpha, \beta,\gamma) = \nonumber  \\[5pt]
&&  \frac{\log \Biggl( 
\ch \sqrt{\frac{1}{4}\alpha}\, \ch \sqrt{\frac{1}{4}\gamma} 
- \frac{\beta}{\sqrt{\alpha\gamma}}\, \sh \sqrt{\frac{1}{4}\alpha}\, \sh \sqrt{\frac{1}{4}\gamma} 
+ \sqrt{\bigl(\ch \sqrt{\frac{1}{4}\alpha}\, \ch \sqrt{\frac{1}{4}\gamma} 
- \frac{\beta}{\sqrt{\alpha\gamma}}\, \sh \sqrt{\frac{1}{4}\alpha}\, \sh \sqrt{\frac{1}{4}\gamma}
\bigr)^2 -1}\Biggr)}
{
\frac
{\sqrt{\frac{1}{4}\alpha} \sqrt{\frac{1}{4}\gamma} } 
{\sh\sqrt{\frac{1}{4}\alpha}\ \sh\sqrt{\frac{1}{4}\gamma} }
\sqrt{
\bigl(\ch \sqrt{\frac{1}{4}\alpha}\, \ch \sqrt{\frac{1}{4}\gamma} 
- \frac{\beta}{\sqrt{\alpha\gamma}}\, \sh \sqrt{\frac{1}{4}\alpha}\, \sh \sqrt{\frac{1}{4}\gamma}
\bigr)^2 -1
}},\ \ \ 
\label{BCH_closed2}
\nonumber  \\
\\
&& G(\alpha) = 
\frac{\sqrt{\frac{1}{4}\alpha}\ \ch \sqrt{\frac{1}{4}\alpha}}{\sh \sqrt{\frac{1}{4}\alpha}},
\ \ \ 
G(\gamma) = 
\frac{\sqrt{\frac{1}{4}\gamma}\ \ch \sqrt{\frac{1}{4}\gamma}}{\sh \sqrt{\frac{1}{4}\gamma}},
\label{BCH_closed3}
\end{eqnarray}
where the symbols `$\ch$' and `$\sh$' are abbreviations of `$\cosh$' and `$\sinh$', respectively.
Notice here that, as functions of the momentum operator $P_{AB}$, $F(\alpha, \beta, \gamma)$, $G(\alpha)$ and $G(\gamma)$ take a value of 1, only when $a_{AB}=b_{AB}=c_{AB}=0$.  

Again, it is important to note that the commutation relations (\ref{XXY3}) and (\ref{YYX3}) induce the
following infinite dimensional Lie algebra,
\begin{eqnarray}
X_{l,m,n} &\equiv& \alpha^{l}\beta^{m}\gamma^{n} X,\ \ \ \ \ l,m,n=0,1,2,...\\[5pt]
Y_{l,m,n} &\equiv& \alpha^{l}\beta^{m}\gamma^{n} Y,\ \ \ \ \ l,m,n=0,1,2,...\\[5pt]
Z_{l,m,n} &\equiv& \alpha^{l}\beta^{m}\gamma^{n} [X,Y],\ \ \ \ \ l,m,n=0,1,2,...,\\[5pt]
[X_{l,m,n},Y_{l',m',n'}]&=&Z_{l+l',m+m',n+n'}\\[5pt]
[X_{l,m,n},Z_{l',m',n'}]&=&X_{l+l',m+m'+1,n+n'} + Y_{l+l',m+m',n+n'+1}\\[5pt]
[Y_{l,m,n},Z_{l',m',n'}]&=&-X_{l+l'+1,m+m',n+n'} - Y_{l+l',m+m'+1,n+n'},
\end{eqnarray}
from which we recognize that the corresponding coordinates of the superspace may again inevitably be
infinite dimensional.

\section{How to circumvent the infinite dimensionality}

\indent

The non-linear appearances of $a_{AB}P_{AB}$, $b_{AB}P_{AB}$, or $c_{AB}P_{AB}$
in every multiplication 
result of the super Lie group elements,  as well as the infinite dimesionality of the superspace
may pose a big obstacle to keeping track of the effect of
$a_{AB}$, $b_{AB}$, or $c_{AB}$,  and to providing an appropriate analysis of supertranslation properties in the superspace
at $a_{AB}, b_{AB}, c_{AB} \neq 0$.
In what follows, we consider how to circumvent the above situation.

\subsection{Trivialization of SUSY Algebra: a naive consideration}

\indent

Let us again begin with the starting SUSY algebra (\ref{SUSY_alg}),
\begin{eqnarray}
\{Q_{A},Q_{B}\} &=& P_{AB}. \nonumber
\end{eqnarray}
One of the  intuitive observations is that if we could divide the both hand sides of the above algebra by $P_{AB}$,
we could erase the effect of $P_{AB}$ from the right hand side of the SUSY algebra, or in other words,
trivialize the right hand side of the SUSY algebra,
\begin{eqnarray}
\Bigl\{\frac{Q_{A}}{\sqrt{P_{AB}}},\frac{Q_{B}}{\sqrt{P_{AB}}}\Bigr\} &=& 1.
\end{eqnarray}
If we could start with this type of algebra, 
all the non-linear factors of $a_{AB}P_{AB}$
appeared in the multiplication results of the super Lie group elements
could be reduced to scaling factors simply depending on $a_{AB}$,
such that the occurrence of the infinite dimensional Lie algebra could be circumvented. 
Of course, this consideration is too naive, since the operator ${Q_{A}}/{\sqrt{P_{AB}}}$ 
can not be expressed as a polynomial in terms of $P_{AB}$, thus may not be well-defined. 

\subsection{Deformation of SUSY Algebra by exponentiating the momentum operator}
\label{deform_SUSY}

\indent

Another possibility is to deform the starting SUSY algebra  (\ref{SUSY_alg}) as
\begin{eqnarray}
\{Q_{A},Q_{B}\} &=& g(P_{AB}), \label{QQg}
\end{eqnarray}
where the function $g(P_{AB})$ is supposed to satisfy,
\begin{eqnarray}
g(P_{AB})\rightarrow P_{AB},  \ \ \ \ {\rm{as}}\ \  a_{AB}\rightarrow 0.
\end{eqnarray}
Examples of the candidates for $g(P_{AB})$ can be given as shift (translation) operators 
$g_{\pm}(P_{AB})$ which may be defined as exponentiations of the momentum operator $P_{AB}$, 
\begin{eqnarray}
g_{\pm}(P_{AB}) &=& \frac{\pm i}{n_{AB}}e^{\mp in_{AB}P_{AB}},\ \ \ {\rm resp.,}
\ \ \ \ (A, B:\rm{no\ sum})
\end{eqnarray}
provided that $g_{\pm}(P_{AB})$ acts on any function $\Phi(x)$ in an adoint manner,
\begin{eqnarray}
[g_{\pm}(P_{AB}),\Phi(x)] &=& \frac{\pm i}{n_{AB}} [e^{\mp in_{AB}P_{AB}}, \Phi(x)]\\[5pt]
&=& [P_{AB}, \Phi(x)] + \mathcal{O}(a_{AB}).
\end{eqnarray}
Here, $n_{AB}$ is introduced as a parameter denoting a magnitude of the shift, and is proportional to $a_{AB}$, namely,
$n_{AB}= a_{AB}/r_{a} $, with $r_{a}$ representing a ratio parameter.

Although the above argument does not depend on any particular representation of the momentum operator $P_{AB}$, it is worthwhile to note that if  $P_{AB}$ is expressed as $P_{AB}=-i\partial_{AB}$
in terms of  a derivative operator $\partial_{AB}$ acting on a bosonic coordinate 
$x$
corresponding to $P_{AB}$,  then
 $g_{\pm }(P_{AB})$ turns out to be a backward  or forward difference operator 
in the following sense,
\begin{eqnarray}
[g_{\pm }(P_{AB}),\Phi(x)] &=& \frac{\pm i}{n_{AB}}(e^{\mp n_{AB}\partial_{AB}}\Phi(x) - \Phi(x)e^{\mp n_{AB}\partial_{AB}}) \\
&=& \frac{\pm i}{n_{AB}}(\Phi(x\mp n_{AB})-\Phi(x))e^{\mp n_{AB}\partial_{AB}} \\[5pt]
&\rightarrow& -i\partial_{AB}\Phi(x), \ \ \ \ {\rm{as}} \ \ \ \ a_{AB}\rightarrow 0.
\end{eqnarray}
Based on the above setup, the SUSY algebra (\ref{QQg}) expressed in terms of the shift operator $g_{+}(P_{AB})$, 
\begin{eqnarray}
\{Q_{A},Q_{B}\} &=& g_{+}(P_{AB})\ =\ \frac{+ i}{n_{AB}}e^{- in_{AB}P_{AB}}, \label{QQg+}
\end{eqnarray}
can be trivialized as, 
\begin{eqnarray}
\{e^{+ia_{A}P_{AB}}Q_{A},e^{+ia_{B}P_{AB}}Q_{B}\} &=& \frac{+ i}{n_{AB}}, \label{QQg+_triv}
\label{triv_Q_alg}
\end{eqnarray}
if we appropriately introduce
shift parameters $a_{A}$ and $a_{B}$ for the supercharges $Q_{A}$ and $Q_{B}$
which satisfy
$ a_{A}+a_{B} = + n_{AB} = +a_{AB}/r_{a}$.
Alternatively, the SUSY algebra (\ref{QQg}) expressed in terms of the shift operator $g_{-}(P_{AB})$,
\begin{eqnarray}
\{Q_{A},Q_{B}\} &=& g_{-}(P_{AB})\ =\ \frac{- i}{n_{AB}}e^{+ in_{AB}P_{AB}}, \label{QQg-}
\end{eqnarray}
can be trivialized as
\begin{eqnarray}
\{e^{+ia_{A}P_{AB}}Q_{A},e^{+ia_{B}P_{AB}}Q_{B}\} &=& \frac{- i}{n_{AB}}, \label{QQg-_triv}
\end{eqnarray}
if the shift parameters $a_{A}$ and $a_{B}$ satisfy
$ a_{A}+a_{B} = - n_{AB}= -a_{AB}/r_{a}$.
In the above context, the conditions 
\begin{eqnarray}
a_{A}+a_{B} &=& +n_{AB}, \ \ \ {\rm for} \ g_{+}(P_{AB}), \label{Triv1}\\[5pt]
a_{A}+a_{B} &=& -n_{AB}, \ \ \ {\rm for} \ g_{-}(P_{AB}), \label{Triv2}
\end{eqnarray}
may be called as trivialization conditions or exponential deformation conditions.

Now, the question is whether there is any SUSY algebra other than the toy model which satisfies the deformation conditions (\ref{Triv1}) and (\ref{Triv2}). 
In a series of papers \cite{DKKN1, DKKN2, DKKN3, DKNS}, extensive attempts have been made to realize exact supersymmetry w.r.t. all the supercharges on a lattice. These formulations introduced a certain type of non-commutativity between $\theta_{A}$ and $x$ in such a way to satisfy the Leibniz rule on a lattice. We now recognize that the above deformation conditions  (\ref{Triv1}) and (\ref{Triv2}) can be identified wtih the lattice Leibniz rule conditions proposed in the above papers.
By utilizing the knowledge of these formulations, we know that $N=D=2$, $N=4,\ D=3$, $N=D=4$ and $N=4, D=5$ 
Dirac-K\"ahler
Twsited SUSY Algebra  (also referred to as Marcus B-type Twisted SUSY Algebra or Geometric Langlands Twisted SUSY Algebra) can satisfy the above conditions 
(\ref{Triv1}) and (\ref{Triv2}).

Before looking at these specific SUSY algebra in detail in the next subsections, 
we would like to make some comments here. 
First, after the trivialization, a careful treatment is needed for the anti-commutators in the left hand sides of (\ref{QQg+_triv}) and (\ref{QQg-_triv}).
As we will see later on, these anti-commutators should be treated as 
shifted anti-commutators or ``link anti-commutators''.

Second, one should notice that introducing the shift operators  $g_{\pm}(P_{AB})$ may cause a complexification of the formulation. For instance, if we begin with a manifest hermitian SUSY algebra (\ref{SUSY_alg}) and manifest hermitian non(anti)commutative parameters (\ref{NAC})
by imposing $Q_{A}^{\dagger} = Q_{A}$, $Q_{B}^{\dagger} = Q_{B}$, $\theta_{A}^{\dagger} = \theta_{A}$,
$\theta_{B}^{\dagger} = \theta_{B}$, we have for supercharges after trivializaion,
\begin{eqnarray}
(e^{+ia_{A}P_{AB}}Q_{A})^{\dagger} &\neq& (e^{+ia_{A}P_{AB}}Q_{A}).
\end{eqnarray}
Also notice that  $g_{+}(P_{AB})$ and $g_{-}(P_{AB})$ are not hermitian by themselves but hermitian conjugate with each other.

Third, 
before promoting the momentum operator $P_{AB}$ to the shift operators $g_{\pm}(P_{AB})$,
the algebra (\ref{SUSY_alg}) is invariant under 
a scale transformation ${\bf{D}}$,
\begin{eqnarray}
[{\bf{D}}, Q_{A}] &=& \frac{1}{2}Q_{A}, \ \ \ 
[{\bf{D}}, Q_{B}] \ =\  \frac{1}{2}Q_{B}, \ \ \ 
[{\bf{D}}, P_{AB}] \ =\ P_{AB}, 
\end{eqnarray}
where we assign the scaling weights $\frac{1}{2}, \frac{1}{2}, 1$ to $Q_{A}, Q_{B}, P_{AB}$, respectively.
Once we promote the  momentum operator $P_{AB}$ to the shift operators $g_{\pm}(P_{AB})$,
the algebra (\ref{QQg+}) is not invariant under the scale transformation, since we introduced a particular length $n_{AB}$ which is a lattice constant.
However, if we consider the following transformation,
\begin{eqnarray}
[{\bf{D}}, Q_{A}] &=& d_{A} Q_{A}, \ \ 
[{\bf{D}}, Q_{B}] \ =\  d_{B} Q_{B}, \label{DQ_cond}
\end{eqnarray}
and operate finite transformation elements of ${\bf{D}}$ on both sides of (\ref{QQg+}), 
we have for $g_{+}(P_{AB})$ and $g_{-}(P_{AB})$, respectively, we have,
\begin{eqnarray}
e^{d_{A}+d_{B} } \{Q_{A},Q_{B}\} &=& 
\frac{+ i}{n_{AB}}\,
e^{{\bf{D}}} 
e^{- in_{AB}P_{AB}}
e^{-{\bf{D}}} ,\\[5pt]
e^{d_{A}+d_{B} } \{Q_{A},Q_{B}\} &=& 
\frac{- i}{n_{AB}}\,
e^{{\bf{D}}} 
e^{+ in_{AB}P_{AB}}
e^{-{\bf{D}}} .
\end{eqnarray}
Using (\ref{QQg+}) again on the left hand sides, 
we obtain,
\begin{eqnarray}
e^{d_{A}+d_{B} }\, e^{- in_{AB}P_{AB}} &=& 
e^{{\bf{D}}}\, e^{- in_{AB}P_{AB}}\, e^{-{\bf{D}}}, \\[5pt]
e^{d_{A}+d_{B} }\, e^{+ in_{AB}P_{AB}} &=& 
e^{{\bf{D}}}\, e^{+ in_{AB}P_{AB}}\, e^{-{\bf{D}}},
\end{eqnarray}
or equivalently,
\begin{eqnarray} 
e^{{\bf{D}}}\, e^{- in_{AB}P_{AB}} &=& 
e^{d_{A}+d_{B} }\, e^{- in_{AB}P_{AB}}\,
e^{{\bf{D}}}, \ \ \ for\ \ g_{+}(P_{AB})  \label{rel5-1}\\[5pt]
e^{{\bf{D}}}\, e^{+ in_{AB}P_{AB}} &=& 
e^{d_{A}+d_{B} }\, e^{+ in_{AB}P_{AB}}\,
e^{{\bf{D}}},   \ \ \ for\ \ g_{-}(P_{AB}) \label{rel5-2}
\end{eqnarray}
which can be regarded as Weyl - `tHooft algebra or group commutator relations 
between $P_{AB}$ and ${\bf{D}}$,
whose degree of non-commutativity is expressed as $d_{A}+d_{B}$.
Sufficient conditions for these relations (\ref{rel5-1}) and  (\ref{rel5-2}) to be satisfied 
can be expressed as
\begin{eqnarray}
[{\bf{D}},P_{AB}] &=& \frac{+ i}{n_{AB}}(d_{A}+d_{B}),  \ \ \ {\rm for}\ \ 
g_{+}(P_{AB}), \label{rel5-3}  \\[5pt]
[{\bf{D}},P_{AB}] &=& \frac{- i}{n_{AB}}(d_{A}+d_{B}),  \ \ \ {\rm for}\ \ 
g_{-}(P_{AB}). \label{rel5-4} 
\end{eqnarray}
If we take $d_{A}$ as $a_{A}$ and $d_{B}$ as $a_{B}$ satisfying (\ref{Triv1}) and (\ref{Triv2}), 
the relations (\ref{rel5-1}) and  (\ref{rel5-2}) 
can be expressed in a unified manner as
\begin{eqnarray}
[{\bf{D}}, P_{AB}] &=& +i, \label{CCR}
\end{eqnarray}
which obviously implies that the symmetry generator $\bf{D}$ may serve as a position operator $x$ corresponding to $P_{AB}$.
The relations (\ref{DQ_cond}) can now be understood as
non-commutativities between 
$Q_{A}$ and $x$, and between $Q_{B}$ and $x$ in the following sense,
\begin{eqnarray}
[x, Q_{A}] &=& a_{A}Q_{A}, \ \ \ \ \ [x, Q_{B}] \ =\ a_{B}Q_{B}. \label{NAC_Qx}
\end{eqnarray}
Conversely, if we start from the relations (\ref{NAC_Qx}) as assumptions
and take commutators of $x$
with both hands side of (\ref{QQg+}) and (\ref{QQg-}),  then
the trivialization conditions (\ref{Triv1}) and (\ref{Triv2}) are naturally derived as a consequence.
The above relations indicate that after promoting the momentum operator $P_{AB}$
to the shift operators $g_{\pm}(P_{AB})$, the position operator $x$ turns out to be a symmetry generator  of the deformed SUSY algebra (\ref{QQg+}) and (\ref{QQg-}).

Another important observation is that
if we anticipate exemplary differential expressions of the supercharges $O_{A}$ and  $Q_{B}$ in the case of
$a_{A}+a_{B}=+n_{AB}$, 
\begin{eqnarray}
Q_{A} &=& \frac{\partial}{\partial\theta_{A}} - \frac{i}{2n_{AB}}\theta_{B}
e^{-in_{AB}P_{AB}},\ \ \ \ \
Q_{B} \ =\  \frac{\partial}{\partial\theta_{B}} - \frac{i}{2n_{AB}}\theta_{A}
e^{-in_{AB}P_{AB}},
\end{eqnarray}
and those for the case of $a_{A}+a_{B}=-n_{AB}$,
\begin{eqnarray}
Q_{A} &=& \frac{\partial}{\partial\theta_{A}}+\frac{i}{2n_{AB}}\theta_{B}
e^{+in_{AB}P_{AB}},\ \ \ \ \
Q_{B} \ =\  \frac{\partial}{\partial\theta_{B}}+\frac{i}{2n_{AB}}\theta_{A}
e^{+in_{AB}P_{AB}},
\end{eqnarray}
it is natural to admit non-commutativities between 
$\frac{\partial}{\partial\theta_{A}}$ and $x$, and between $\frac{\partial}{\partial\theta_{B}}$ and $x$,
\begin{eqnarray}
[x, \frac{\partial}{\partial\theta_{A}}] &=& a_{A}\frac{\partial}{\partial\theta_{A}}, 
\ \ \ \ \ [x, \frac{\partial}{\partial\theta_{B}}] \ =\ a_{B}\frac{\partial}{\partial\theta_{B}}, \label{NAC_del_theta_x}
\end{eqnarray}
as well as 
non-commutativities between 
$\theta_{A}$ and $x$, and between $\theta_{B}$ and $x$,
\begin{eqnarray}
[x, \theta_{A}] &=& -a_{A}\theta_{A}, 
\ \ \ \ \ [x, \theta_{B}] \ =\ -a_{B}\theta_{B}. \label{NAC_ax}
\end{eqnarray}
Reminding that the non-commutativities between $Q_{A,B}$ and $x$ (\ref{NAC_Qx})
can be traced back to the existence of the shift operators in the right-hand sides of 
(\ref{QQg+}) and (\ref{QQg-}), one may observe that 
the  non-commutativities between $\theta_{A,B}$ and $x$ (\ref{NAC_ax}) naturally implies
an existence of inverse shift operators 
in the right-hand side of (\ref{NAC}), namely,
\begin{eqnarray}
\{\theta_{A},\theta_{B}\} &\sim & a_{AB}\, e^{+in_{AB}P_{AB}}, \ \ \ {\rm{for}}\ \  a_{A}+a_{B}=+n_{AB} , \\[5pt]
\{\theta_{A},\theta_{B}\} &\sim & a_{AB}\,e^{-in_{AB}P_{AB}}, \ \ \ {\rm{for}}\ \  a_{A}+a_{B}=-n_{AB},
\end{eqnarray}
which contain the contributions of the momentum $P_{AB}$  at $\mathcal{O}(a_{AB}^{2})$. This means that the contributions of $P_{AB}$ to the non(anti)commutativity
between  $\theta_{A}$'s are negligible for small $a_{AB}$.
Although, in principle, there are several possible choices of coefficients in front of the 
exponentials, we suppose the following Ansatz in this article,
\begin{eqnarray}
\{\theta_{A},\theta_{B}\} &= & r_{a}\, g_{+}^{-1}(P_{AB})\ =\ 
-i\, a_{AB}\, e^{+in_{AB}P_{AB}}, \ \ \ {\rm{for}}\ \  a_{A}+a_{B}=+n_{AB} , \label{NACe1}\\[5pt]
\{\theta_{A},\theta_{B}\} &= & r_{a}\, g_{-}^{-1}(P_{AB})\ =\ 
+i\, a_{AB}\,e^{-in_{AB}P_{AB}}, \ \ \ {\rm{for}}\ \  a_{A}+a_{B}=-n_{AB}, \label{NACe2}
\end{eqnarray}
where $r_{a}$ again represents the ratio parameter, $r_{a}=a_{AB}/n_{AB}$, introduced above.

Under the above Ansatz, the situation becomes much simpler. If we start from the deformed SUSY algebra (\ref{QQg+}) and (\ref{QQg-}) together with the corresponding non(anti)commutative Grassmann relations (\ref{NACe1}) and (\ref{NACe2}) instead of (\ref{SUSY_alg}) and (\ref{NAC}),
the element $a_{AB}P_{AB}$ non-linearly appeared in
the previous section is now simply reduced to the momentum independent ratio parameter 
$r_{a}=a_{AB} / n_{AB}$ 
for both cases of $g_{+}(P_{AB})$ and $g_{-}(P_{AB})$,
\begin{eqnarray}
a_{AB}P_{AB} &\rightarrow & r_{a}\, g_{+}^{-1}(P_{AB})\, g_{+}(P_{AB}) \ = \ r_{a} \\[5pt]
a_{AB}P_{AB} &\rightarrow & r_{a}\, g_{-}^{-1}(P_{AB})\, g_{-}(P_{AB}) \ = \ r_{a}.
\end{eqnarray}
This means that the non-linear factors appeared in the Baker-Campbell-Haussdorff
formulae in the previous section are now governed by the ratio parameter $r_{a}$.
If we repeat the similar argument, the elements $b_{AB}P_{AB}$ and $c_{AB}P_{AB}$ appeared 
in the previous section can also be reduced to the momentum independent ratio factors
 $r_{b}$ and $r_{c}$, respectively, where $r_{b}=b_{AB}/n_{AB}$ and  $r_{c}=c_{AB}/n_{AB}$.

In a similar manner, one may introduce non-commutativities between 
$\frac{\partial}{\partial \theta_{A}}$, $\frac{\partial}{\partial \theta_{B}}$ and
$x$ (\ref{NAC_del_theta_x}) as,
\begin{eqnarray}
\bigl\{
\frac{\partial}{\partial \theta_{A}}, \frac{\partial}{\partial \theta_{B}} 
\bigr\}
&=& t_{a}\, g_{+} (P_{AB}),\ \ \ \ {\rm for}\ \ \  n_{A} + n_{B} =+n_{AB}, \\[5pt]
\bigl\{
\frac{\partial}{\partial \theta_{A}}, \frac{\partial}{\partial \theta_{B}} 
\bigr\}
&=& t_{a}\, g_{-} (P_{AB}),\ \ \ \ {\rm for}\ \ \  n_{A} + n_{B} =-n_{AB}, 
\end{eqnarray}
where $t_{a}$ denotes another ratio parameter whose value may be determined 
in accordance with the ratio parameter $r_{a}$.
One may observe that the above non(anti)commutative properties contain a sort of 
redundancy in the sense that the inverse of the translation operators, $g^{-1}_{\pm}$,
is encoded in the anti-commutation relation between $\theta_{A}$ and $\theta_{B}$, 
while 
the anti-commutation relation between 
$\frac{\partial}{\partial \theta_{A}}$ and $\frac{\partial}{\partial \theta_{B}}$
already represents the translation operators, $g_{\pm}$.
We will come back to this point later on.

\section{Dirac-K\"ahler twisted SUSY Algebra with exponential deformation
in various dimensions}
\label{DK_alg_various}

\subsection{$N=D=2$ Twisted SUSY Algebra}
\label{subsectionN=D=2}

\indent

As a first explicit example of the trivializable SUSY algebra, in the following, we briefly look at the $N=2$ Twisted SUSY Algebra in two dimensional spacetime. For more detailed explanations and discussions regarding topologically twisted supersymmetrirs, the readers may refer to 
\cite{Witten1,DKKN1,DKKN2, KT,KKU}. See also \cite{AKKM} for the analysis with a central charge.

The $N=D=2$ untwisted super-Poincar\'e algebra 
in Euclidean continuum spacetime can be given by,
\begin{eqnarray}
\{Q_{\alpha i},\overline{Q}_{j \beta}\} &=&
2\delta_{ij}(\gamma^{\mu})_{\alpha\beta}P_{\mu},\\[2pt]
[J,Q_{\alpha i}] &=& \frac{i}{2}(\gamma^{5})_{\alpha}^{\ \beta}Q_{\beta i},\\[2pt]
[R,Q_{\alpha i}] &=& \frac{i}{2}(\gamma^{5})_{i}^{\ j}Q_{\alpha j},\\[2pt]
[J,P_{\mu}] &=&i\epsilon_{\mu\nu}P_{\nu}, \\[4pt]
[R,P_{\mu}] &=& [P_{\mu},Q_{\alpha i}]\ =\ [P_{\mu},P_{\nu}]\ =\
[J,R]\ =\ 0,
\end{eqnarray}
where $\alpha,\beta$ and $i,j$ denote spinor and 
$N=2$ internal indices, respectively.
$J$ and $R$ denote generators for $SO(2)$ spacetime and $SO(2)$
internal rotations.
In $N=2$ in 2-dimensional 
spacetime,
$\overline{Q}_{i\alpha}$ is subject to
\begin{eqnarray}
\overline{Q}_{i\alpha} = (C^{-1}Q^{T}C)_{i\alpha} = Q_{\alpha i}
\end{eqnarray}
where $C$ represents charge conjugation matrix which can be taken as
$C=1$.

One can perform twisting of the above algebra through
the definition of twisted Lorentz generator $J'$ as
\begin{eqnarray}
J' \equiv J+R, .
\end{eqnarray}
which corresponds to a generator of the diagonal subgroup $(SO(2)_{J}\times SO(2)_{R})$.
Corresponding super-Poincar\'e algebra can be 
most naturally described by the following
Dirac-K\"ahler expanded supercharges,
\begin{eqnarray}
Q_{\alpha i} = (\mathbf{1}Q + \gamma^{\mu}Q_{\mu} + \gamma^{5}\tilde{Q})_{\alpha i},
\label{DK2d}
\end{eqnarray}
as
\begin{eqnarray}
\{Q,Q_{\mu}\} &=& P_{\mu}, \label{N=2tsusy1} \\[2pt]
\{\tilde{Q},Q_{\mu}\} &=& -\epsilon_{\mu\nu}P_{\nu},\label{N=2tsusy2}\\[4pt]
Q^{2}\ =\ \tilde{Q}^{2} &=& \{Q,\tilde{Q}\}\ =\ \{Q_{\mu},Q_{\nu}\}\ =\ 0,\label{N=2tsusy3} 
\end{eqnarray}
where $\epsilon_{\mu\nu}$ is the two dimensional totally anti-symmetric
tensor with $\epsilon_{12}=+1$.
With the above Dirac-K\"ahler twisted basis,
$J$ and $R$ rotations are expressed as
\begin{eqnarray}
[J,Q] &=& -\frac{i}{2}\tilde{Q}, \hspace{20pt}
[J,\tilde{Q}]\ =\ +\frac{i}{2}Q, \label{2DJQ1} \hspace{20pt} 
[J,Q_{\mu}] \ =\  +\frac{i}{2}\epsilon_{\mu\nu}Q_{\nu}, \hspace{20pt} \label{2DJQ2}
\\[2pt] 
[R,Q] &=& +\frac{i}{2}\tilde{Q}, \hspace{20pt} 
[R,\tilde{Q}]\ =\ -\frac{i}{2}Q, \label{2DRQ1} \hspace{20pt} 
[R,Q_{\mu}] \ =\  +\frac{i}{2}\epsilon_{\mu\nu}Q_{\nu}, \hspace{30pt} \label{2DRQ2}
\end{eqnarray}
and for $J'=J+R$,
\begin{eqnarray}
[J',Q] &=& [J',\tilde{Q}]\ =\ 0, \hspace{20pt} 
[J',Q_{\mu}] \ =\ i\epsilon_{\mu\nu}Q_{\mu}.
\end{eqnarray}
Notice that the twisted supercharges $(Q,Q_{\mu},\tilde{Q})$
transform as scalar, vector and (pseudo) scalar, respectively,
under the twisted Lorentz rotation $J'$.

The exponential deformation of  (\ref{N=2tsusy1}) and (\ref{N=2tsusy2}) can be done,
through promoting  the momentum operator  $P_{\mu}$ in the right-hand sides 
to the shift operators, as,
\begin{eqnarray}
\{Q,Q_{\mu}\} &=& g_{+}(P_{\mu})\ =\ \frac{+i}{n}e^{-i\vec{n}_{\mu}\cdot \vec{P}}, \label{N=2tsusy1_tr} \\[2pt]
\{\tilde{Q},Q_{\mu}\} &=& -\epsilon_{\mu\nu}g_{-}({P_{\nu}})\ =\
- \epsilon_{\mu\nu}\frac{-i}{n}e^{+i\vec{n}_{\nu}\cdot \vec{P}}.\label{N=2tsusy2_tr}
\end{eqnarray}
Here we take a uniform lattice 
where $\vec{n}_{\mu}$ represents a vector whose components are given by $(\vec{n}_{\mu})_{\rho} = n\, \delta_{\mu\rho}$,
in other words, $\vec{n}_{1}=(n,0)$, $\vec{n}_{2}=(0,n)$,
where $n$ is representing  a length of the vector $\vec{n}_{\mu}$  
which is  a lattice constant for the two dimensional spacetime.

Commutation relations between $\vec{x}$ and $(Q,Q_{\mu},\tilde{Q})$,
and the deformation conditions associated with (\ref{N=2tsusy1_tr}) and (\ref{N=2tsusy1_tr})
can be expressed, in terms of shift vectors $(\vec{a}, \vec{a}_{\mu}, \vec{\tilde{a}})$
for the supercharges $(Q,Q_{\mu},\tilde{Q})$, as
\begin{eqnarray}
[\vec{x}, Q] &=& \vec{a}\,  Q,\ \ \ 
[\vec{x}, Q_{\mu}] \ =\  \vec{a}_{\mu}\,  Q_{\mu},\ \ \ 
[\vec{x}, \tilde{Q}] \ =\ \vec{\tilde{a}}\,  \tilde{Q}, \ \ \ \ ({\rm \mu : no\ sum})
\label{comm_Q_x}
\end{eqnarray}
and 
\begin{eqnarray}
\vec{a} + \vec{a}_{\mu}  &=& +\vec{n}_{\mu}, \label{triv_cond1}\\[5pt]
\vec{\tilde{a}} + \vec{a}_{\mu}  &=& -|\epsilon_{\mu\nu}|\vec{n}_{\nu}. \label{triv_cond2}
\end{eqnarray}
Note here that the conditions (\ref{triv_cond1}) (\ref{triv_cond2}) have the following generic solutions
with one vector arbitrariness,
\begin{eqnarray}
\vec{a} &=& (arbitrary),\ \ \ \ \vec{a}_{\mu} \ = \ +\vec{n}_{\mu} - \vec{a}, \ \ \ \ 
\vec{\tilde{a}} \ = \ -\vec{n}_{1} -\vec{n}_{2} + \vec{a},
\end{eqnarray}
and one can also see, from the conditions (\ref{triv_cond1}) (\ref{triv_cond2}),
that the total sum of the shifts associated with the supercharges turn out to be 0, namely,
\begin{eqnarray}
\vec{a} + \vec{a}_{1} + \vec{a}_{2} +\vec{\tilde{a}} &=& \vec{0}.
\end{eqnarray}
One of the particular choices of the shift vectors can be expressed as,
\begin{eqnarray}
\frac{\vec{a}}{n} &= &(+\frac{1}{2},+\frac{1}{2}),
\ \ \ \frac{\vec{a}_{1}}{n} \ = \ (+\frac{1}{2},-\frac{1}{2}), \ \ \ 
\frac{\vec{a}_{2}}{n} \ = \ (-\frac{1}{2},+\frac{1}{2}),
\ \ \ \frac{\vec{\tilde{a}}}{n}\ =\  (-\frac{1}{2},-\frac{1}{2}),
\end{eqnarray}
which may be called as a symmetric choice. Another particular choice can be expressed as,
\begin{eqnarray}
\frac{\vec{a}}{n} &= &(0,0),
\ \ \ \frac{\vec{a}_{1}}{n} \ = \ (+1,0), 
\ \ \ \frac{\vec{a}_{2}}{n} \ = \ (0,+1), 
\ \ \ \frac{\vec{\tilde{a}}}{n}\ =\  (-1,-1),
\end{eqnarray}
which may be called as an asymmetric choice.

The commutation relations (\ref{comm_Q_x}) imply commutation relations 
between $\vec{x}$ and $N=D=2$ twisted Grassmann coordinates
$(\theta,\theta_{\mu},\tilde{\theta})$,
\begin{eqnarray}
[\vec{x}, \theta] &=& -\vec{a}\,  \theta,\ \ \ 
[\vec{x}, \theta_{\mu}] \ =\  -\vec{a}_{\mu}\,  \theta_{\mu},\ \ \ 
[\vec{x}, \tilde{\theta}] \ =\ -\vec{\tilde{a}}\,  \tilde{\theta}, \ \ \ \ ({\rm \mu : no\ sum})
\label{comm_theta_x}
\end{eqnarray}
from which we may admit anti-commutation relations among the Grassmann coordinates,
\begin{eqnarray}
\{\theta,\theta_{\mu}\} &=& r_{a}^{(2)}\, g_{+}^{-1}(P_{\mu})\ =
\ - i\, r_{a}^{(2)}\, n\, e^{+i\vec{n}_{\mu}\cdot \vec{P}}, \label{N=2tsusy1_tr_theta} \\[5pt]
\{\tilde{\theta},\theta_{\mu}\} &=& -r_{a}^{(2)}\, \epsilon_{\mu\nu}g_{-}^{-1}({P_{\nu}})\ =\
-i r_{a}^{(2)}\, n\, \epsilon_{\mu\nu} e^{-i\vec{n}_{\nu}\cdot \vec{P}}, \label{N=2tsusy2_tr_theta}
\end{eqnarray}
where $r_{a}^{(2)}$ representing a ratio parameter 
explained in the previous subsection, applied in the case of $N=D=2$. 
Correspondingly, anti-commutation relations among the derivatives of Grassmann coordinates may be introduced as,
\begin{eqnarray}
\bigl\{ \frac{\partial}{\partial\theta}, \frac{\partial}{\partial\theta_{\mu}}
 \bigr\} &=& t_{a}^{(2)} g_{+}(P_{\mu}), \label{N=2tsusy1_tr_dtheta} \\[5pt]
\bigl\{ \frac{\partial}{\partial\tilde{\theta}},
\frac{\partial}{\partial\theta_{\mu}}
 \bigr\} &=& -t_{a}^{(2)} \epsilon_{\mu\nu}g_{-}(P_{\nu}), \label{N=2tsusy2_tr_dtheta}
\end{eqnarray}
where $t_{a}^{(2)}$ representing another ratio parameter 
explained in the previous subsection, applied in the case of $N=D=2$. 


\subsection{$N=4\ D=3$ Twisted SUSY Algebra}
\label{subsectionN=4D=4}

\indent

As a second explicit example of the trivializable SUSY algebra, we pick up $N=4$ Twisted SUSY Algebra in three dimensional spacetime \cite{DKKN3, Nagata1}.
$N=4\ D=3$ untwisted SUSY algebra can be given by, 
\begin{eqnarray}
&&\hspace{20pt} \{Q_{\alpha i},\overline{Q}_{j\beta}\} \ =\ 
2\delta_{ij}(\gamma_{\mu})_{\alpha\beta}P_{\mu}, \label{N=4D=3SUSYalgebra}\\[2pt]
[J_{\mu},Q_{\alpha i}] &=& +\frac{1}{2}(\gamma_{\mu})_{\alpha\beta}Q_{\beta i},
\hspace{30pt}
[J_{\mu},\overline{Q}_{i\alpha}] \ =\ -\frac{1}{2}
\overline{Q}_{i\beta}(\gamma_{\mu})_{\beta\alpha},\\[0pt]
[R_{\mu},Q_{\alpha i}] &=& -\frac{1}{2}Q_{\alpha j}(\gamma_{\mu})_{ji},
\hspace{30pt}
[R_{\mu},\overline{Q}_{i\alpha}] \ =\ +\frac{1}{2}
(\gamma_{\mu})_{ij}\overline{Q}_{j\alpha},\\[2pt]
[J_{\mu},P_{\nu}] &=& -i\epsilon_{\mu\nu\rho}P_{\rho},
\hspace{20pt}
[J_{\mu},J_{\nu}] \ =\ -i\epsilon_{\mu\nu\rho}J_{\rho},
\hspace{20pt}
[R_{\mu},R_{\nu}] \ =\ -i\epsilon_{\mu\nu\rho}R_{\rho}, \\[4pt]
[R_{\mu},P_{\nu}] &=&
[P_{\mu},P_{\nu}] \ =\ 
[J_{\mu},R_{\nu}] \ =\ 0,
\end{eqnarray}
where the gamma matrices $\gamma_{\mu}$
can be taken as the Pauli matrices,
$\gamma^{\mu}(\mu=1,2,3)\equiv (\sigma^{1},\sigma^{2},\sigma^{3})$.
The conjugate supercharge $\overline{Q}_{i\alpha}$ 
can be taken as the complex conjugation
of $Q_{\alpha i}$, $\overline{Q}_{i\alpha}=Q^{*}_{\alpha i}$.
The $J_{\mu}$ and $R_{\mu}\ (\mu =1,2,3)$ are the generators of 
$SO(3)_{E}\simeq SU(2)_{E}$ Euclidean Lorentz rotations 
and $SO(3)_{R}\simeq SU(2)_{R}$ internal rotations, respectively.

As in the case of $N=D=2$,
the twisting procedure can be performed
by taking the diagonal subgroup of 
the Lorentz rotations and the internal rotations.
Here in the case of $N=4\ D=3$, we take the diagonal subgroup
$(SO(3)_{E}\times SO(3)_{R})_{diag}$ 
whose covering group is $(SU(2)_{E}\times SU(2)_{R})_{diag}$.
This corresponds to 
introducing the twisted Lorentz generator $J'_{\mu}$ as
a diagonal sum of $J_{\mu}$ and $R_{\mu}$,
$J'_{\mu} \equiv J_{\mu} + R_{\mu}$.
Since, after the twisting, the Lorentz index $\alpha$ and internal index $i$
are rotated on the same footing,
the resulting algebra is most naturally expressed in terms of the following
Dirac-K\"ahler expansion of the supercharges,
\begin{eqnarray}
Q_{\alpha i} &=& (\mathbf{1}Q+\gamma_{\mu}Q_{\mu})_{\alpha i}, 
\hspace{30pt}
\overline{Q}_{i\alpha} \ =\ 
(\mathbf{1}\overline{Q}+\gamma_{\mu}\overline{Q}_{\mu})_{i\alpha},
\end{eqnarray}
where $\bf{1}$ represents the two-by-two unit matrix.
The coefficients 
$(Q,\overline{Q}_{\mu},Q_{\mu},\overline{Q})$
are the $N=4\ D=3$ twisted supercharges.
After the expansions, the original SUSY algebra (\ref{N=4D=3SUSYalgebra})
can be expressed, in terms of the twisted basis, as,
\begin{eqnarray}
\{Q,\overline{Q}_{\mu}\} &=& P_{\mu}, \label{N=4D=3_SUSYalgebra1}\\[2pt]
\{Q_{\mu},\overline{Q}_{\nu}\} & =& -i\epsilon_{\mu\nu\rho}P_{\rho},
\label{N=4D=3_SUSYalgebra2}\\[2pt]
\{\overline{Q},Q_{\mu}\} & =& P_{\mu}, \label{N=4D=3_SUSYalgebra3}\\[2pt]
\{others\} & =& 0,\qquad \label{N=4D=3twistedSUSYalgebra4}
\end{eqnarray}
where $\epsilon_{\mu\nu\rho}$ is the three dimensional totally anti-symmetric
tensor with $\epsilon_{123}=+1$.
The Lorentz and the internal rotations 
of the supercharges are re-expressed on the
twisted basis,
\begin{eqnarray}
[J_{\mu},Q] &=& +\frac{1}{2}Q_{\mu}, 
\hspace{20pt}
[J_{\mu},Q_{\nu}] \ =\ -\frac{i}{2}\epsilon_{\mu\nu\rho}Q_{\rho}
+\frac{1}{2}\delta_{\mu\nu}Q, \label{3DsJ1} \\[0pt]
[J_{\mu},\overline{Q}] &=& -\frac{1}{2}\overline{Q}_{\mu}, 
\hspace{20pt}
[J_{\mu},\overline{Q}_{\nu}] \ =\ -\frac{i}{2}\epsilon_{\mu\nu\rho}\overline{Q}_{\rho}
-\frac{1}{2}\delta_{\mu\nu}\overline{Q},\label{3DsJ2} \\[0pt]
[R_{\mu},Q] &=& -\frac{1}{2}Q_{\mu}, 
\hspace{20pt}
[R_{\mu},Q_{\nu}] \ =\ -\frac{i}{2}\epsilon_{\mu\nu\rho}Q_{\rho}
-\frac{1}{2}\delta_{\mu\nu}Q, \label{3DsJ3} \\[0pt]
[R_{\mu},\overline{Q}] &=& +\frac{1}{2}\overline{Q}_{\mu},
\hspace{20pt}
[R_{\mu},\overline{Q}_{\nu}] \ =\ -\frac{i}{2}\epsilon_{\mu\nu\rho}\overline{Q}_{\rho}
+\frac{1}{2}\delta_{\mu\nu}\overline{Q}. \label{3DsJ4}
\end{eqnarray}
Notice that $(Q,\overline{Q})$ and $(Q_{\mu},\overline{Q}_{\mu})$ transform as
scalars and vectors
under $(SO(3)_{E}\times SO(3)_{R})_{diag}$,
respectively.
Namely, under the twisted Lorentz generator 
$J'_{\mu}=J_{\mu}+R_{\mu}$ they transform as
\begin{eqnarray}
[J'_{\mu},Q] &=& [J'_{\mu},\overline{Q}]\ = \ 0,\hspace{20pt} 
[J'_{\mu},Q_{\nu}] \ =\ -i\epsilon_{\mu\nu\rho}Q_{\rho},\hspace{20pt} 
[J'_{\mu},\overline{Q}_{\nu}] \ =\ -i\epsilon_{\mu\nu\rho}\overline{Q}_{\rho}.
\qquad
\end{eqnarray}

Historically, studies of the $N=4\ D=3$ twisted SYM in the continuum spacetime
have been given in \cite{BT,GM} with the classification of
two inequivalent topological twists which are called the super BF type (A-type)
and the Blau-Thompson type (B-type).
Both of these two types give the same SUSY algebra 
(\ref{N=4D=3_SUSYalgebra1})-(\ref{N=4D=3_SUSYalgebra3}),
although the corresponding gauge multiplets are different with each other.
We note that the lattice Dirac-K\"ahler twisted 
SYM multiplet given in \cite{DKKN3}
is categorized as the B-type twist as explicitly discussed in \cite{Nagata1}.. 

The exponential deformation of  (\ref{N=4D=3_SUSYalgebra1})-(\ref{N=4D=3_SUSYalgebra3}) can be done,
through promoting  the momentum operator  $P_{\mu}$ in the right-hand sides 
to the shift operators, as,
\begin{eqnarray}
\{Q,\overline{Q}_{\mu}\} &=& g_{+}(P_{\mu}) \ =\ \frac{+i}{n}e^{-i\vec{n}_{\mu}\cdot \vec{P}},
\label{N=4D=3_SUSYalgebra_lat1}\\[2pt]
\{Q_{\mu},\overline{Q}_{\nu}\} & =& -i\epsilon_{\mu\nu\rho}
g_{-}(P_{\rho}) \ =\ -i \epsilon_{\mu\nu\rho} \frac{-i}{n} e^{+i\vec{n}_{\rho}\cdot \vec{P}},
\label{N=4D=3_SUSYalgebra_lat2}\\[2pt]
\{\overline{Q},Q_{\mu}\} & =& g_{+}(P_{\mu}) \ =\ \frac{+i}{n}e^{-i\vec{n}_{\mu}\cdot \vec{P}},
\label{N=4D=3_SUSYalgebra_lat3}
\end{eqnarray}
where $\vec{n}_{\mu}$ again denotes a vector whose components are given by $(\vec{n}_{\mu})_{\rho} = n\, \delta_{\mu\rho}$,
in other words, $\vec{n}_{1}=(n,0,0)$, $\vec{n}_{2}=(0,n,0)$, $\vec{n}_{3}=(0,0,n)$,
while $n$ is representing  a length of the vector $\vec{n}_{\mu}$  
which is  a lattice constant for the three dimensional spacetime.

Commutation relations between $\vec{x}$ and $(Q,\overline{Q}_{\mu},Q_{\mu},\overline{Q})$,
and the deformation conditions associated with (\ref{N=4D=3_SUSYalgebra_lat1})- (\ref{N=4D=3_SUSYalgebra_lat3})
can be expressed, in terms of shift vectors 
$(\vec{a}, \vec{\overline{a}}_{\mu}, \vec{a}_{\mu}, \vec{\overline{a}})$
for the supercharges $(Q,\overline{Q}_{\mu},Q_{\mu},\overline{Q})$, as
\begin{eqnarray}
[\vec{x}, Q] &=& \vec{a}\,  Q,\ \ \ 
[\vec{x}, \overline{Q}_{\mu}] \ =\  \vec{\overline{a}}_{\mu}\,  \overline{Q}_{\mu},\ \ \ [\vec{x}, Q_{\mu}] \ =\  \vec{a}_{\mu}\,  Q_{\mu},\ \ \ 
[\vec{x}, \overline{Q}] \ =\ \vec{\overline{a}}\,  \overline{Q}, \ \ \ \ ({\rm \mu : no\ sum})\ \ \
\label{comm_Q_x_3d}
\end{eqnarray}
and 
\begin{eqnarray}
\vec{a} + \vec{\overline{a}}_{\mu}  &=& +\vec{n}_{\mu}, \label{triv_cond1_3d}\\[5pt]
\vec{a}_{\mu} + \vec{\overline{a}}_{\nu}  &=& -|\epsilon_{\mu\nu\rho}|
\vec{n}_{\rho}, \label{triv_cond2_3d}\\[5pt]
\vec{\overline{a}} + \vec{a}_{\mu}  &=& +\vec{n}_{\mu}. \label{triv_cond3_3d}
\end{eqnarray}
Note here that the conditions (\ref{triv_cond1_3d})-(\ref{triv_cond3_3d}) have the following generic solutions
with one vector arbitrariness,
\begin{eqnarray}
\vec{a} &=& (arbitrary),\ \ \ \ 
\vec{\overline{a}}_{\mu} \ = \ +\vec{n}_{\mu} - \vec{a}, \\[5pt] \ \ \ \ 
\vec{{a}}_{\mu} & = & -\sum_{\lambda\neq \mu} \vec{n}_{\lambda} + \vec{a}, \ \ \ \ 
\vec{\overline{a}} \ = \ +\sum_{\lambda = 1}^{3}
\vec{n}_{\lambda} - \vec{a},
\end{eqnarray}
and one can also see, from the conditions (\ref{triv_cond1_3d})-(\ref{triv_cond3_3d}),
that the total sum of the shifts associated with the supercharges turn out to be 0, namely,
\begin{eqnarray}
\vec{a} + \sum_{\mu = 1}^{3} \vec{\overline{a}}_{\mu}
+ \sum_{\mu = 1}^{3} \vec{{a}}_{\mu}
+\vec{\overline{a}} &=& \vec{0}.
\end{eqnarray}
One of the particular choices of the shift vectors can be expressed as,
\begin{eqnarray}
\frac{\vec{a}}{n} &= &(+\frac{1}{2},+\frac{1}{2},+\frac{1}{2}),
\ \ \ \ \ \frac{\vec{\overline{a}}}{n} \ = \ (+\frac{1}{2},+\frac{1}{2},+\frac{1}{2}), \\[5pt] 
\frac{\vec{a}_{1}}{n} & = & (+\frac{1}{2},-\frac{1}{2},-\frac{1}{2}),
\ \ \ \ \  \frac{\vec{\overline{a}}_{1}}{n}\ =\  (+\frac{1}{2},-\frac{1}{2},-\frac{1}{2}), \\[5pt]
\frac{\vec{a}_{2}}{n} & = & (-\frac{1}{2},+\frac{1}{2},-\frac{1}{2}),
\ \ \ \ \ \frac{\vec{\overline{a}}_{2}}{n}\ =\  (-\frac{1}{2},+\frac{1}{2},-\frac{1}{2}), \\[5pt]
\frac{\vec{a}_{3}}{n} & = & (-\frac{1}{2},-\frac{1}{2},+\frac{1}{2}),
\ \ \ \ \ \frac{\vec{\overline{a}}_{3}}{n}\ =\  (-\frac{1}{2},-\frac{1}{2},+\frac{1}{2}), 
\end{eqnarray}
which may be called as a symmetric choice. Another particular choice can be expressed as,
\begin{eqnarray}
\frac{\vec{a}}{n} &= & (0,0,0),
\ \ \ \ \ \frac{\vec{\overline{a}}}{n} \ = \ (+1,+1,+1), \\[5pt] 
\frac{\vec{a}_{1}}{n} & = & (0,-1,-1),
\ \ \ \ \  \frac{\vec{\overline{a}}_{1}}{n}\ =\  (+1,0,0), \\[5pt]
\frac{\vec{a}_{2}}{n} & = & (-1,0,-1),
\ \ \ \ \ \frac{\vec{\overline{a}}_{2}}{n}\ =\  (0,+1,0), \\[5pt]
\frac{\vec{a}_{3}}{n} & = & (-1,-1,0),
\ \ \ \ \ \frac{\vec{\overline{a}}_{3}}{n}\ =\  (0,0,+1), 
\end{eqnarray}
which may be called as an asymmetric choice. 

The commutation relations (\ref{comm_Q_x_3d}) imply commutation relations 
between $\vec{x}$ and $N=4\ D=3$ twisted Grassmann coordinates
$(\theta,\overline{\theta}_{\mu}, \theta_{\mu},\overline{\theta})$,
\begin{eqnarray}
[\vec{x}, \theta] &=& -\vec{a}\,  \theta,\ \ \ 
[\vec{x}, \overline{\theta}_{\mu}] \ =\  -\vec{\overline{a}}_{\mu}\,  \overline{\theta}_{\mu},\ \ \ 
[\vec{x}, \theta_{\mu}] \ =\  -\vec{a}_{\mu}\,  \theta_{\mu},\ \ \ 
[\vec{x}, \overline{\theta}] \ =\ -\vec{\overline{a}}\,  \overline{\theta}, \ \ \ ({\rm \mu : no\ sum}) \ \ \ \ \
\label{comm_theta_x_3d}
\end{eqnarray}
from which we may admit anti-commutation relations among the Grassmann coordinates,
\begin{eqnarray}
\{\theta,\overline{\theta}_{\mu}\} &=& r_{a}^{(3)} g_{+}^{-1}(P_{\mu}) \ =\ 
-i\, r_{a}^{(3)}\, n\, 
e^{+i\vec{n}_{\mu}\cdot \vec{P}},
\label{N=4D=3_SUSYalgebra_lat1_theta}\\[5pt]
\{\theta_{\mu},\overline{\theta}_{\nu}\} & =& +ir_{a}^{(3)} \epsilon_{\mu\nu\rho}
g_{-}^{-1}(P_{\rho}) \ =\ -r_{a}^{(3)}\, n\, \epsilon_{\mu\nu\rho} 
e^{-i\vec{n}_{\rho}\cdot \vec{P}},
\label{N=4D=3_SUSYalgebra_lat2_theta}\\[5pt]
\{\overline{\theta},\theta_{\mu}\} & =& r_{a}^{(3)} g_{+}^{-1}(P_{\mu}) \ =\ 
-i\, r_{a}^{(3)}\, n\, e^{+i\vec{n}_{\mu}\cdot \vec{P}},
\label{N=4D=3_SUSYalgebra_lat3_theta}
\end{eqnarray}
where $r_{a}^{(3)}$ representing a ratio parameter in the case of $N=4\ D=3$. 
Correspondingly, anti-commutation relations among the derivatives of Grassmann coordinates may be introduced as,
\begin{eqnarray}
\bigl\{ \frac{\partial}{\partial\theta}, \frac{\partial}{\partial\overline{\theta}_{\mu}}
 \bigr\} &=& t_{a}^{(3)} g_{+}(P_{\mu}), \\[5pt]
\bigl\{ \frac{\partial}{\partial\theta_{\mu}},
\frac{\partial}{\partial\overline{\theta}_{\nu}}
 \bigr\} &=& +i t_{a}^{(3)} \epsilon_{\mu\nu\rho}g_{-}(P_{\rho}), \\[5pt]
\bigl\{ \frac{\partial}{\partial\overline{\theta}}, \frac{\partial}{\partial\theta_{\mu}}
 \bigr\} &=& t_{a}^{(3)} g_{+}(P_{\mu})
\end{eqnarray}
where $t_{a}^{(3)}$ representing another ratio parameter 
explained in the previous subsection, applied in the case of $N=4\ D=3$.

\subsection{$N=D=4$ Twisted SUSY Algebra}
\label{subsectionN=D=4}

\indent

As a third explicit example of the trivializable SUSY algebra, we pick up $N=4$ Twisted SUSY Algebra in four dimensional spacetime. Full analysis of the algebra can be found  in \cite{KKM}.
See also \cite{Saito1} for the formulation with a central charge.

We start from the following $N=4$ extended SUSY algebra in 4-dimensional Euclidean spacetime,
\begin{eqnarray}
\{Q_{\alpha i},\overline{Q}_{j\beta}\} &=& 2\delta_{ij}(\gamma^{\mu})_{\alpha\beta}P_{\mu},
\label{N=D=4susy}
\end{eqnarray}
where $\overline{Q}_{j\beta}$ is defined by the following conjugation of $Q_{\alpha i}$,
\begin{eqnarray}
\overline{Q}_{i\alpha} &=& (C^{-1}Q^{T}C)_{i\alpha},
\end{eqnarray}
with charge conjugation matrix $C$ which satisfies
\begin{eqnarray}
\gamma_{\mu}^{T} &=& C\gamma_{\mu}C^{-1},\ \ \ \ \
C^{T} \ =\ -C. 
\end{eqnarray}
The remaining part of extended super Poincare algebra can be defined as
\begin{eqnarray}
[J_{\mu\nu},Q_{\alpha i}] &=& -\frac{i}{2}(\gamma_{\mu\nu})_{\alpha\beta}Q_{\beta i},\\[0pt]
[R_{\mu\nu},Q_{\alpha i}] &=& +\frac{i}{2}Q_{\alpha j}(\gamma_{\mu\nu})_{ji},\\[2pt]
[P_{\mu},Q_{\alpha i}] &=& 0,\\[2pt]
[J_{\mu\nu},P_{\rho}] &=& -i\delta_{\mu\nu\rho\sigma}P_{\sigma},\\[2pt]
[R_{\mu\nu},P_{\rho}] &=& 0,\\[2pt]
[P_{\mu},P_{\nu}] &=& 0,\\[2pt]
[J_{\mu\nu},J_{\rho\sigma}] &=& -i(\delta_{\mu\rho}J_{\nu\sigma}-\delta_{\nu\rho}J_{\mu\sigma}
-\delta_{\mu\sigma}J_{\nu\rho}+\delta_{\nu\sigma}J_{\mu\rho}),\\[2pt]
[R_{\mu\nu},R_{\rho\sigma}] &=& -i(\delta_{\mu\rho}R_{\nu\sigma}-\delta_{\nu\rho}R_{\mu\sigma}
-\delta_{\mu\sigma}R_{\nu\rho}+\delta_{\nu\sigma}R_{\mu\rho}),\\[2pt]
[J_{\mu\nu},R_{\rho\sigma}] &=& 0,
\end{eqnarray}
where $J_{\mu\nu}$ and $R_{\mu\nu}$ denote $SO(4)$ Lorentz and $SO(4) \subset SU(4)$ of 
$N=4$ internal rotation generators, respectively. 

Dirac-K\"ahler twisting procedure can be done by
introducing ``twisted" Lorentz generator $J'_{\mu\nu}$ as
\begin{eqnarray}
J'_{\mu\nu} &=& J_{\mu\nu} + R_{\mu\nu}.
\end{eqnarray}
The corresponding twisted super Poincare algebra is most naturally
described through the following 
Dirac-K\"ahler expansion of supercharge $Q_{\alpha i}$,
\begin{eqnarray}
Q_{\alpha i} &=& \frac{1}{\sqrt{2}}(\mathbf{1}Q + \gamma^{\mu}Q_{\mu}
+\frac{1}{2}\gamma^{\mu\nu}Q_{\mu\nu} + \tilde{\gamma^{\mu}}\tilde{Q}_{\mu}
+ \gamma^{5}\tilde{Q})_{\alpha i}, \label{4DQ}
\end{eqnarray}
where $\gamma^{\mu\nu}\equiv\frac{1}{2}[\gamma^{\mu},\gamma^{\nu}]$,
$\tilde{\gamma}^{\mu}\equiv \gamma^{\mu}\gamma^{5}$, 
$\gamma^{5}\equiv \gamma^{1}\gamma^{2}\gamma^{3}\gamma^{4}$ 
with $\{\gamma^\mu,\gamma^\nu\}=2\delta^{\mu\nu}$.

SUSY algebra of (\ref{N=D=4susy}) can be written by using the above expansion as
\begin{eqnarray}
\{Q,Q_{\mu}\} &=& P_{\mu},\label{N=D=4tsusy1}\\[2pt]
\{Q_{\rho\sigma},Q_{\mu}\} &=& -\delta_{\rho\sigma\mu\nu}P_{\nu},
\label{N=D=4tsusy2}\\[2pt]
\{Q_{\rho\sigma},\tilde{Q}_{\mu}\} &=& +\epsilon_{\rho\sigma\mu\nu}P_{\nu},
\label{N=D=4tsusy3}\\[2pt]
\{\tilde{Q},\tilde{Q}_{\mu}\} &=& P_{\mu},\label{N=D=4tsusy4}\\[2pt]
\{others\} &=& 0, \label{N=D=4tsusy5}
\end{eqnarray}
and also for the remaining part of the algebra,
\begin{eqnarray}
[J_{\mu\nu},Q] &=& +\frac{i}{2}Q_{\mu\nu}, \label{Lorentz4D1}\\[2pt]
[J_{\mu\nu},Q_{\rho}]  &=&  -\frac{i}{2}
\delta_{\mu\nu\rho\sigma}Q_{\sigma}
-\frac{i}{2}\epsilon_{\mu\nu\rho\sigma}\tilde{Q}_{\sigma},\\[2pt]
[J_{\mu\nu},Q_{\rho\sigma}]&=&-\frac{i}{2}\delta_{\mu\nu\rho\sigma}Q
+\frac{i}{2}\epsilon_{\mu\nu\rho\sigma}\tilde{Q}
+\frac{i}{2}\delta_{\mu\nu\rho\lambda}Q_{\sigma\lambda}
-\frac{i}{2}\delta_{\mu\nu\sigma\lambda}Q_{\rho\lambda},\\[2pt]
[J_{\mu\nu},\tilde{Q}_{\rho}]&=&-\frac{i}{2}\epsilon_{\mu\nu\rho\sigma}
Q_{\sigma}-\frac{i}{2}\delta_{\mu\nu\rho\sigma}\tilde{Q}_{\sigma}\\[2pt]
[J_{\mu\nu},\tilde{Q}] &=& -\frac{i}{4}\epsilon_{\mu\nu\rho\sigma}
Q_{\rho\sigma}, \label{Lorentz4D5} \\[7pt]
[J_{\mu\nu},P_{\rho}]&=&-i\delta_{\mu\nu\rho\sigma}P_{\sigma},\\[10pt]
[J_{\mu\nu},J_{\rho\sigma}]&=&
-i(\delta_{\mu\rho}J_{\nu\sigma}-\delta_{\nu\rho}J_{\mu\sigma}
-\delta_{\mu\sigma}J_{\nu\rho}+\delta_{\nu\sigma}J_{\mu\rho}),
\end{eqnarray}
\begin{eqnarray}
[R_{\mu\nu},Q] &=& -\frac{i}{2}Q_{\mu\nu},\label{R4D1}\\[2pt]
[R_{\mu\nu},Q_{\rho}]  &=&  -\frac{i}{2}
\delta_{\mu\nu\rho\sigma}Q_{\sigma}
+\frac{i}{2}\epsilon_{\mu\nu\rho\sigma}\tilde{Q}_{\sigma},\\[2pt]
[R_{\mu\nu},Q_{\rho\sigma}]&=&+\frac{i}{2}\delta_{\mu\nu\rho\sigma}Q
-\frac{i}{2}\epsilon_{\mu\nu\rho\sigma}\tilde{Q}
+\frac{i}{2}\delta_{\mu\nu\rho\lambda}Q_{\sigma\lambda}
-\frac{i}{2}\delta_{\mu\nu\sigma\lambda}Q_{\rho\lambda},\\[2pt]
[R_{\mu\nu},\tilde{Q}_{\rho}]&=&+\frac{i}{2}\epsilon_{\mu\nu\rho\sigma}
Q_{\sigma}
-\frac{i}{2}\delta_{\mu\nu\rho\sigma}\tilde{Q}_{\sigma}\\[2pt]
[R_{\mu\nu},\tilde{Q}] &=& +\frac{i}{4}\epsilon_{\mu\nu\rho\sigma}
Q_{\rho\sigma},\label{R4D5}\\[7pt]
[R_{\mu\nu},P_{\rho}]&=&0,\\[5pt]
[R_{\mu\nu},R_{\rho\sigma}]&=&
-i(\delta_{\mu\rho}R_{\nu\sigma}-\delta_{\nu\rho}R_{\mu\sigma}
-\delta_{\mu\sigma}R_{\nu\rho}+\delta_{\nu\sigma}R_{\mu\rho}),\\[5pt]
[J_{\mu\nu},R_{\rho\sigma}]&=& [P_{\mu},P_{\nu}]\ =\ 0,
\end{eqnarray} 
where $\delta_{\mu\nu\rho\sigma}\equiv\delta_{\mu\rho}\delta_{\nu\sigma}
-\delta_{\mu\sigma}\delta_{\nu\rho}$. 
Notice that under twisted Lorentz generator $J'_{\mu}$,
each component of Dirac-K\"ahler twisted supercharge
transforms as
\begin{eqnarray}
[J'_{\mu\nu},Q] &=& 0,\\[2pt]
[J'_{\mu\nu},Q_{\rho}] &=& -i\delta_{\mu\nu\rho\sigma}Q_{\sigma},\\[2pt]
[J'_{\mu\nu},Q_{\rho\sigma}]&=& i\delta_{\mu\nu\rho\lambda}Q_{\sigma\lambda}
-i\delta_{\mu\nu\sigma\lambda}Q_{\rho\lambda},\\[2pt]
[J'_{\mu\nu},\tilde{Q}_{\rho}] &=& -i\delta_{\mu\nu\rho\sigma}\tilde{Q}_{\sigma},\\[2pt]
[J'_{\mu\nu},\tilde{Q}] &=& 0,
\end{eqnarray}
which means each component $(Q,Q_{\mu},Q_{\mu\nu},\tilde{Q}_{\mu},\tilde{Q})$ behaves
as (scalar, vector, 2nd rank tensor, (pseudo)vector, (pseudo)scalar), respectively, 
in the space where $J'_{\mu\nu}$ is rotation generators.

Historically, there have been known three different types of $N=4$ twisting in four dimensions.
Here we briefly review these three types.
The twisting for $N=4$
comes from the observation that  
$SU(4)$ splits into its regular subgroups
such as 
\begin{equation}
SU(4)\supset SO(4) \simeq SU(2)_1\otimes SU(2)_2.
\end{equation}
Accordingly, the representation $\mathbf{4}$
of $SU(4)_I$ splits into the following
three types,
\begin{equation}
\begin{split}
 (i) \ \ & \mathbf{4}\rightarrow 
(\mathbf{2},\mathbf{1})\oplus
(\mathbf{2},\mathbf{1}) \\
 (i\hspace{-1pt}i) \ \ & \mathbf{4}\rightarrow 
(\mathbf{2},\mathbf{1})\oplus
(\mathbf{1},\mathbf{2})\\ 
 (i\hspace{-1pt}i\hspace{-1pt}i) 
\ \ & \mathbf{4}\rightarrow 
(\mathbf{2},\mathbf{1})\oplus
(\mathbf{1},\mathbf{1})\oplus
(\mathbf{1},\mathbf{1})
 \end{split}
\end{equation}
with the indices $(SU(2)_1,SU(2)_2)$.

Twisting procedure for $N=4$ in general 
is to take $SU(2)'_L$ and $SU(2)'_R$
as diagonal subgroups of 
$SU(2)_L\otimes SU(2)_1$ and
$SU(2)_R\otimes SU(2)_2$ ,
respectively.
Accordingly, the untwisted supercharges 
$Q_{\alpha i}(\mathbf{2},\mathbf{1},\mathbf{4}) $ and
$\overline{Q}_{j \beta}(\mathbf{1},\mathbf{2},\mathbf{4}) $
may be split into the following twisted representations.

For case $(i)$,
\begin{equation}
\begin{split}
&
(\mathbf{2},\mathbf{1},\mathbf{4}) 
\rightarrow
2(\mathbf{1},\mathbf{1})
\oplus
2(\mathbf{3},\mathbf{1}) \\
&
(\mathbf{1},\mathbf{2},\mathbf{4}) 
\rightarrow
2(\mathbf{2},\mathbf{2})
\end{split}
\end{equation}

For case $(i\hspace{-1pt}i)$,
\begin{equation}
\begin{split}
&
(\mathbf{2},\mathbf{1},\mathbf{4}) 
\rightarrow
(\mathbf{1},\mathbf{1})
\oplus
(\mathbf{3},\mathbf{1})
\oplus
(\mathbf{2},\mathbf{2}) \\
&
(\mathbf{1},\mathbf{2},\mathbf{4}) 
\rightarrow
(\mathbf{2},\mathbf{2})
\oplus
(\mathbf{1},\mathbf{1})
\oplus
(\mathbf{1},\mathbf{3}).
\end{split}
\label{fermion}
\end{equation}

For case $(i\hspace{-1pt}i\hspace{-1pt}i)$,
\begin{equation}
\begin{split}
&
(\mathbf{2},\mathbf{1},\mathbf{4}) 
\rightarrow
(\mathbf{1},\mathbf{1})
\oplus
(\mathbf{3},\mathbf{1})
\oplus
2(\mathbf{2},\mathbf{1}) \\
&
(\mathbf{1},\mathbf{2},\mathbf{4}) 
\rightarrow
(\mathbf{2},\mathbf{2})
\oplus
2(\mathbf{1},\mathbf{2}).
\end{split}
\end{equation}

The first case (\textit{i}) is called A-type which
 was first introduced by Yamron \cite{Yamron}
 and studied by Vafa and Witten in a strong coupling test of S-duality \cite{Vafa-Witten}.
The second case (\textit {ii}) 
is called B-type which
was first briefly introduced at the end of \cite{Yamron},
and extensively 
studied by Marcus in \cite{Marcus} as another type of TQFT
by means of complexifying the gauge fields.
 The third case (\textit{iii}) was also first introduced in \cite{Yamron}, and is
also called half-twisted theory \cite{Lozano}. 
As for the studies of $N=D=4$ twisting, see also 
\cite{D=N=4_etc}.
Note that as pointed out in \cite{Kapustin_Witten}, the B-type twisting has been playing an important role in the studies of geometric Langlands correspondence, therefore, it is 
also called as  geometric Langlands twist or GL-twist.

We can see from the above representations that
the above explained  Dirac-K\"ahler twisting can be identified with 
 the second case (\textit {ii}), which is B-type twist or GL-twist, 
 since it includes 
 (scalar$(\mathbf{1},\mathbf{1})$, vector$(\mathbf{2},\mathbf{2})$, 
 2nd rank tensor$(\mathbf{3},\mathbf{1})\oplus(\mathbf{1},\mathbf{3})$, 
 (pseudo)vector$(\mathbf{2},\mathbf{2})$, (pseudo)scalar$(\mathbf{1},\mathbf{1})$)
 as the twisted representation of the supercharges.

The exponential deformation of  (\ref{N=D=4tsusy1})-(\ref{N=D=4tsusy4}) can be done,
through promoting  the momentum operator  $P_{\mu}$ in the right-hand sides 
to the shift operators, as,
\begin{eqnarray}
\{Q,Q_{\mu}\} &=& g_{+}(P_{\mu}) \ =\ \frac{+i}{n}e^{-i\vec{n}_{\mu}\cdot \vec{P}}
\label{N=D=4tsusy1_lat} \\[2pt] 
\{Q_{\rho\sigma},Q_{\mu}\} &=& -\delta_{\rho\sigma\mu\nu}
g_{-}(P_{\nu}) \ =\ -\delta_{\rho\sigma\mu\nu}
\frac{-i}{n}e^{+i\vec{n}_{\nu}\cdot \vec{P}}  \label{N=D=4tsusy2_lat}\\[2pt]
\{Q_{\rho\sigma},\tilde{Q}_{\mu}\} &=& +\epsilon_{\rho\sigma\mu\nu}
g_{+}(P_{\nu}) \ =\  +\epsilon_{\rho\sigma\mu\nu}
\frac{+i}{n}e^{-i\vec{n}_{\nu}\cdot \vec{P}} \label{N=D=4tsusy3_lat} \\[2pt]
\{\tilde{Q},\tilde{Q}_{\mu}\} &=& g_{-}(P_{\mu}) \ =\ \frac{-i}{n}e^{+i\vec{n}_{\mu}\cdot \vec{P}}
\label{N=D=4tsusy4_lat}
\end{eqnarray}
where $\vec{n}_{\mu}$ again denotes a vector whose components are given by $(\vec{n}_{\mu})_{\rho} = n\, \delta_{\mu\rho}$,
in other words, $\vec{n}_{1}=(n,0,0,0)$, $\vec{n}_{2}=(0,n,0,0)$, $\vec{n}_{3}=(0,0,n,0)$,
$\vec{n}_{4}=(0,0,0,n)$,
while $n$ is representing  a length of the vector $\vec{n}_{\mu}$  
which is  a lattice constant for the four dimensional spacetime.

Commutation relations between $\vec{x}$ and 
 $(Q,Q_{\mu},Q_{\mu\nu},\tilde{Q}_{\mu},\tilde{Q})$,
and the deformation conditions associated with 
(\ref{N=D=4tsusy1_lat})-(\ref{N=D=4tsusy4_lat})
can be expressed, in terms of shift vectors 
$(\vec{a}, \vec{a}_{\mu}, \vec{a}_{\mu\nu}, \vec{\tilde{a}}_{\mu},  \vec{\tilde{a}})$
for the supercharges  $(Q,Q_{\mu},Q_{\mu\nu},\tilde{Q}_{\mu},\tilde{Q})$, as
\begin{eqnarray}
[\vec{x}, Q] \ =\ \vec{a}\,  Q,\ \ \ 
[\vec{x}, Q_{\mu}] \ =\  \vec{a}_{\mu}\,  Q_{\mu},\ \ \ 
[\vec{x}, Q_{\mu\nu}] \ =\  \vec{a}_{\mu\nu}\,  Q_{\mu\nu},\ \ \ 
[\vec{x}, \tilde{Q}_{\mu}] & =&  \vec{\tilde{a}}_{\mu}\,  \tilde{Q}_{\mu},\ \ \ 
[\vec{x}, \tilde{Q}] \ =\ \vec{\tilde{a}}\,  \tilde{Q}, \ \ \ \ \nonumber \\[2pt]
&& ({\rm \mu, \nu : no\ sum})\ \ \
\label{comm_Q_x_4d}
\end{eqnarray}
and 
\begin{eqnarray}
\vec{a} + \vec{a}_{\mu}  &=& +\vec{n}_{\mu}, 
\label{triv_cond1_4d}\\[5pt]
\vec{a}_{\rho\sigma} + \vec{a}_{\mu} &=& -|\delta_{\rho\sigma\mu\nu}| \vec{n}_{\nu} 
\ \ \ for\ \  \rho = \mu \ \ or\ \  \sigma = \mu
\label{triv_cond2_4d}\\[5pt]
\vec{a}_{\rho\sigma} + \vec{\tilde{a}}_{\mu} &=& +|\epsilon_{\rho\sigma\mu\nu}| \vec{n}_{\nu} 
\ \ \ for\ \  \rho, \sigma,  \mu : {\it all\ different\ each\ other}
\label{triv_cond3_4d} \\[5pt]
\vec{\tilde{a}} + \vec{\tilde{a}}_{\mu}  &=& -\vec{n}_{\mu}. \label{triv_cond3_4d}
\end{eqnarray}

As in the case of $N=D=2$ and $N=4\ D=3$, there exists one parameter arbitrariness
for the solution of the shift vectors $(\vec{a}, \vec{a}_{\mu}, \vec{a}_{\mu\nu}, \vec{\tilde{a}}_{\mu},  \vec{\tilde{a}})$.
One of the typical choice for the shift vectors, the symmetric choice, is given as,
\begin{eqnarray}
\frac{\vec{a}}{n} &=& (+\frac{1}{2},+\frac{1}{2},+\frac{1}{2},+\frac{1}{2}), \hspace{20pt}
\frac{\vec{a}_{1}}{n}\ =\ (+\frac{1}{2},-\frac{1}{2},-\frac{1}{2},-\frac{1}{2}), \\[2pt]
\frac{\vec{a}_{12}}{n} &=& (-\frac{1}{2},-\frac{1}{2},+\frac{1}{2},+\frac{1}{2}), \hspace{20pt}
\frac{\vec{a}_{2}}{n}\ =\ (-\frac{1}{2},+\frac{1}{2},-\frac{1}{2},-\frac{1}{2}), \\[2pt] 
\frac{\vec{a}_{13}}{n} &=& (-\frac{1}{2},+\frac{1}{2},-\frac{1}{2},+\frac{1}{2}), \hspace{20pt}
\frac{\vec{a}_{3}}{n}\ =\ (-\frac{1}{2},-\frac{1}{2},+\frac{1}{2},-\frac{1}{2}), \\[2pt] 
\frac{\vec{a}_{14}}{n} &=& (-\frac{1}{2},+\frac{1}{2},+\frac{1}{2},-\frac{1}{2}), \hspace{20pt}
\frac{\vec{a}_{4}}{n}\ =\ (-\frac{1}{2},-\frac{1}{2},-\frac{1}{2},+\frac{1}{2}), \\[2pt] 
\frac{\vec{a}_{23}}{n} &=& (+\frac{1}{2},-\frac{1}{2},-\frac{1}{2},+\frac{1}{2}), \hspace{20pt}
\frac{\vec{\tilde{a}}_{4}}{n}\ =\ (+\frac{1}{2},+\frac{1}{2},+\frac{1}{2},-\frac{1}{2}), \\[2pt] 
\frac{\vec{a}_{24}}{n} &=& (+\frac{1}{2},-\frac{1}{2},+\frac{1}{2},-\frac{1}{2}), \hspace{20pt}
\frac{\vec{\tilde{a}}_{3}}{n}\ =\ (+\frac{1}{2},+\frac{1}{2},-\frac{1}{2},+\frac{1}{2}), \\[2pt] 
\frac{\vec{a}_{34}}{n} &=& (+\frac{1}{2},+\frac{1}{2},-\frac{1}{2},-\frac{1}{2}), \hspace{20pt}
\frac{\vec{\tilde{a}}_{2}}{n}\ =\ (+\frac{1}{2},-\frac{1}{2},+\frac{1}{2},+\frac{1}{2}), \\[2pt]
\frac{\vec{\tilde{a}}}{n} &=& (-\frac{1}{2},-\frac{1}{2},-\frac{1}{2},-\frac{1}{2}), \hspace{20pt}
\frac{\vec{\tilde{a}}_{1}}{n}\ =\ (-\frac{1}{2},+\frac{1}{2},+\frac{1}{2},+\frac{1}{2}), 
\end{eqnarray}
while the other typical choice, the asymmetric choice, is given as,
\begin{eqnarray}
\frac{\vec{a}}{n} &=& (0,0,0,0), \hspace{57pt}
\frac{\vec{a}_{1}}{n}\ =\ (+1,0,0,0), \\[2pt]
\frac{\vec{a}_{12}}{n} &=& (-1,-1,0,0), \hspace{38pt}
\frac{\vec{a}_{2}}{n}\ =\ (0,+1,0,0), \\[2pt] 
\frac{\vec{a}_{13}}{n} &=& (-1,0,-1,0), \hspace{38pt}
\frac{\vec{a}_{3}}{n}\ =\ (0,0,+1,0), \\[2pt] 
\frac{\vec{a}_{14}}{n} &=& (-1,0,0,-1), \hspace{38pt}
\frac{\vec{a}_{4}}{n}\ =\ (0,0,0,+1), \\[2pt] 
\frac{\vec{a}_{23}}{n} &=& (0,-1,-1,0), \hspace{38pt}
\frac{\vec{\tilde{a}}_{4}}{n}\ =\ (+1,+1,+1,0), \\[2pt] 
\frac{\vec{a}_{24}}{n} &=& (0,-1,0,-1), \hspace{38pt}
\frac{\vec{\tilde{a}}_{3}}{n}\ =\ (+1,+1,0,+1), \\[2pt] 
\frac{\vec{a}_{34}}{n} &=& (0,0,-1,-1), \hspace{38pt}
\frac{\vec{\tilde{a}}_{2}}{n}\ =\ (+1,0,+1,+1), \\[2pt]
\frac{\vec{\tilde{a}}}{n} &=& (-1,-1,-1,-1), \hspace{20pt}
\frac{\vec{\tilde{a}}_{1}}{n}\ =\ (0,+1,+1,+1).
\end{eqnarray}

The commutation relations (\ref{comm_Q_x_4d}) imply commutation relations between $\vec{x}$ and 
$N =D=4$ twisted Grassmann coordinates
$(\theta, \theta_{\mu}, \theta_{\mu\nu}, \tilde{\theta}_{\mu}, \tilde{\theta})$,
\begin{eqnarray}
[\vec{x}, \theta] \ =\ -\vec{a}\,  \theta,\ \ \ 
[\vec{x}, \theta_{\mu}] \ =\  -\vec{a}_{\mu}\,  \theta_{\mu},\ \ \ 
[\vec{x}, \theta_{\mu\nu}] \ =\  -\vec{a}_{\mu\nu}\,  \theta_{\mu\nu},\ \ \ 
[\vec{x}, \tilde{\theta}_{\mu}] & =&  -\vec{\tilde{a}}_{\mu}\,  \tilde{\theta}_{\mu},\ \ \ 
[\vec{x}, \tilde{\theta}] \ =\ -\vec{\tilde{a}}\,  \tilde{\theta}, \ \ \ \ \nonumber \\[2pt]
&& ({\rm \mu, \nu : no\ sum})\ \ \
\label{comm_theta_x_4d}
\end{eqnarray}
from which we may admit anti-commutation relations among the Grassmann coordinates,
\begin{eqnarray}
\{\theta,\theta_{\mu}\} &=& r^{(4)}_{a}\, g^{-1}_{+}(P_{\mu}) \ =\ 
-i\, r^{(4)}_{a}\, n\, e^{+i\vec{n}_{\mu}\cdot \vec{P}} 
\label{N=D=4tsusy1_lat_theta} \\[2pt] 
\{\theta_{\rho\sigma},\theta_{\mu}\} &=& - r^{(4)}_{a}\, \delta_{\rho\sigma\mu\nu}
g^{-1}_{-}(P_{\nu}) \ =\ 
- i\, r^{(4)}_{a}\, n\, \delta_{\rho\sigma\mu\nu}\,
e^{-i\vec{n}_{\nu}\cdot \vec{P}}  \label{N=D=4tsusy2_lat_theta}\\[2pt]
\{\theta_{\rho\sigma},\tilde{\theta}_{\mu}\} &=& 
+ r^{(4)}_{a}\, \epsilon_{\rho\sigma\mu\nu}
g^{-1}_{+}(P_{\nu}) \ =\   -i\, r^{(4)}_{a}\, n\,  \epsilon_{\rho\sigma\mu\nu}
e^{+i\vec{n}_{\nu}\cdot \vec{P}} \label{N=D=4tsusy3_lat_theta} \\[2pt]
\{\tilde{\theta},\tilde{\theta}_{\mu}\} &=&r^{(4)}_{a}\,  g^{-1}_{-}(P_{\mu}) 
\ =\ +i\, r^{(4)}_{a}\, n\,
e^{-i\vec{n}_{\mu}\cdot \vec{P}}
\label{N=D=4tsusy4_lat_theta}
\end{eqnarray}
where $r_{a}^{(4)}$ representing a ratio parameter in the case of $N=D=4$.
Correspondingly, anti-commutation relations among the derivatives of Grassmann coordinates may be introduced as,
\begin{eqnarray}
\bigl\{ \frac{\partial}{\partial\theta}, \frac{\partial}{\partial\theta_{\mu}}
 \bigr\} &=& t_{a}^{(4)} g_{+}(P_{\mu}), \\[5pt]
\bigl\{ \frac{\partial}{\partial\theta_{\rho\sigma}},
\frac{\partial}{\partial\theta_{\mu}}
 \bigr\} &=& - t_{a}^{(4)} \delta_{\rho\sigma\mu\nu}g_{-}(P_{\nu}), \\[5pt]
 \bigl\{ \frac{\partial}{\partial\theta_{\rho\sigma}},
\frac{\partial}{\partial\tilde{\theta}_{\mu}}
 \bigr\} &=& + t_{a}^{(4)} \epsilon_{\rho\sigma\mu\nu}g_{+}(P_{\nu}), \\[5pt]
\bigl\{ \frac{\partial}{\partial\tilde{\theta}}, \frac{\partial}{\partial\tilde{\theta}_{\mu}}
 \bigr\} &=& t_{a}^{(4)} g_{-}(P_{\mu})
\end{eqnarray}
where $t_{a}^{(4)}$ representing another ratio parameter 
explained in the previous subsection, applied in the case of $N=D=4$.

\subsection{$N=4\ D=5$ Twisted SUSY Algebra}
\label{subsectionN=4D=5}

\indent

$N=4\ D=5$ Twisted SUSY Algebra can be obtained by a five-dimensional lift-up of the 
$N=D=4$ Twisted SUSY algebra (\ref{N=D=4tsusy1})-(\ref{N=D=4tsusy4}), as follows \cite{DKNS},
\begin{eqnarray}
\{Q,Q_{\mu}\} &=& P_{\mu},\label{N=4D=5tsusy1}\\[2pt]
\{Q_{\rho\sigma},Q_{\mu}\} &=& -\delta_{\rho\sigma\mu\nu}P_{\nu},
\label{N=4D=5tsusy2}\\[2pt]
\{Q_{\rho\sigma},Q_{\mu\nu}\} &=& +\epsilon_{\rho\sigma\mu\nu\lambda}P_{\lambda},
\label{N=4D=5tsusy3}
\end{eqnarray}
where $\delta_{\mu\nu\rho\sigma}\equiv \delta_{\mu\rho}\delta_{\nu\sigma}-\delta_{\mu\sigma}\delta_{\nu\rho}$ and $\epsilon_{\rho\sigma\mu\nu\lambda}$
is a ﬁve dimensional
totally anti-symmetric tensor with $\epsilon_{12345}\equiv +1$, while
$Q_{\mu\nu}$ is a five dimensional totally anti-symmetric tensor 
consisting of 10 components.
If one performs a dimensional reduction with the following identiﬁcations
\begin{eqnarray}
P_{5} &\rightarrow& 0, \ \ \ \ \  Q_{\mu 5}\ \rightarrow\ -\tilde{Q}_{\mu}, 
\ \ \ \ \ \ 
Q_{5} \ \rightarrow\ -\tilde{Q},
\end{eqnarray}
the $N=4\ D=5$ Twisted SUSY Algebra (\ref{N=4D=5tsusy1})-(\ref{N=4D=5tsusy3})
is reduced to  the 
$N=D=4$ Twisted SUSY algebra (\ref{N=D=4tsusy1})-(\ref{N=D=4tsusy4}).
We note that
 $N = 4\ D=5$  Dirac-K\"ahler twisted SUSY
algebra
have been worked out  also in \cite{CGJ}, 
by paying a particular
attention to R symmetries.

The exponential deformation of  (\ref{N=4D=5tsusy1})-(\ref{N=4D=5tsusy3}) can be done, through promoting the momentum
operator $P_{\mu}$ in the right-hand sides to the shift operators, as,
\begin{eqnarray}
\{Q,Q_{\mu}\} &=& 
g_{+}(P_{\mu}) \ =\ \frac{+i}{n}e^{-i\vec{n}_{\mu}\cdot \vec{P}}
\label{N=4D=5tsusy1_lat}\\[2pt]
\{Q_{\rho\sigma},Q_{\mu}\} &=& 
-\delta_{\rho\sigma\mu\nu}\,
g_{-}(P_{\nu}) \ =\ -\delta_{\rho\sigma\mu\nu}
\frac{-i}{n}e^{+i\vec{n}_{\nu}\cdot \vec{P}} 
\label{N=4D=5tsusy2_lat}\\[2pt]
\{Q_{\rho\sigma},Q_{\mu\nu}\} &=& 
+\epsilon_{\rho\sigma\mu\nu\lambda}\,
g_{+}(P_{\lambda}) \ =\
+\epsilon_{\rho\sigma\mu\nu\lambda}
\frac{+i}{n}e^{-i\vec{n}_{\lambda}\cdot \vec{P}} 
\label{N=4D=5tsusy3_lat}
\end{eqnarray}
where $\vec{n}_{\mu}$ again denotes a vector whose components are given by $(\vec{n}_{\mu})_{\rho} = n\, \delta_{\mu\rho}$,
in other words, $\vec{n}_{1}=(n,0,0,0,0)$, $\vec{n}_{2}=(0,n,0,0,0)$, $\vec{n}_{3}=(0,0,n,0,0)$,
$\vec{n}_{4}=(0,0,0,n,0)$, $\vec{n}_{5}=(0,0,0,0,n)$
while $n$ is representing  a length of the vector $\vec{n}_{\mu}$  
which is  a lattice constant for the five dimensional spacetime.

Commutation relations between $\vec{x}$ and 
 $(Q,Q_{\mu},Q_{\mu\nu})$,
and the deformation conditions associated with 
(\ref{N=4D=5tsusy1_lat})-(\ref{N=4D=5tsusy3_lat})
can be expressed, in terms of shift vectors 
$(\vec{a}, \vec{a}_{\mu}, \vec{a}_{\mu\nu})$
for the supercharges  $(Q,Q_{\mu},Q_{\mu\nu})$, as
\begin{eqnarray}
[\vec{x}, Q] \ =\ \vec{a}\,  Q,\ \ \ 
[\vec{x}, Q_{\mu}] \ =\  \vec{a}_{\mu}\,  Q_{\mu},\ \ \ 
[\vec{x}, Q_{\mu\nu}] \ =\  \vec{a}_{\mu\nu}\,  Q_{\mu\nu},\ \ \ 
 ({\rm \mu, \nu : no\ sum})\ \ \
\label{comm_Q_x_5d}
\end{eqnarray}
and 
\begin{eqnarray}
\vec{a} + \vec{a}_{\mu}  &=& +\vec{n}_{\mu}, 
\label{triv_cond1_5d}\\[5pt]
\vec{a}_{\mu\nu} + \vec{a}_{\mu} &=& - \vec{n}_{\nu} 
\label{triv_cond2_5d}\\[5pt]
\vec{a}_{\rho\sigma} + \vec{a}_{\mu\nu} &=& +|\epsilon_{\rho\sigma\mu\nu\lambda}| \vec{n}_{\lambda} 
\ \ \ for\ \  \rho, \sigma,  \mu, \nu : {\it all\ different\ each\ other}.
\label{triv_cond3_5d}
\end{eqnarray}
As shown in \cite{DKNS}, the deformation conditions (\ref{triv_cond1_5d})-(\ref{triv_cond3_5d}) allows only the symmetric choice for the shift vectors,
\begin{eqnarray}
\frac{\vec{a}}{n} &=& (+\frac{1}{2},+\frac{1}{2},+\frac{1}{2},+\frac{1}{2},+\frac{1}{2}), \hspace{20pt}
\frac{\vec{a}_{1}}{n}\ =\ (+\frac{1}{2},-\frac{1}{2},-\frac{1}{2},-\frac{1}{2},-\frac{1}{2}), \\[2pt]
\frac{\vec{a}_{12}}{n} &=& (-\frac{1}{2},-\frac{1}{2},+\frac{1}{2},+\frac{1}{2},+\frac{1}{2}), \hspace{20pt}
\frac{\vec{a}_{2}}{n}\ =\ (-\frac{1}{2},+\frac{1}{2},-\frac{1}{2},-\frac{1}{2},-\frac{1}{2}), \\[2pt] 
\frac{\vec{a}_{13}}{n} &=& (-\frac{1}{2},+\frac{1}{2},-\frac{1}{2},+\frac{1}{2},+\frac{1}{2}), \hspace{20pt}
\frac{\vec{a}_{3}}{n}\ =\ (-\frac{1}{2},-\frac{1}{2},+\frac{1}{2},-\frac{1}{2},-\frac{1}{2}), \\[2pt] 
\frac{\vec{a}_{14}}{n} &=& (-\frac{1}{2},+\frac{1}{2},+\frac{1}{2},-\frac{1}{2},+\frac{1}{2}), \hspace{20pt}
\frac{\vec{a}_{4}}{n}\ =\ (-\frac{1}{2},-\frac{1}{2},-\frac{1}{2},+\frac{1}{2},-\frac{1}{2}), \\[2pt] 
\frac{\vec{a}_{15}}{n} &=& (-\frac{1}{2},+\frac{1}{2},+\frac{1}{2},+\frac{1}{2},-\frac{1}{2}), \hspace{20pt}
\frac{\vec{a}_{5}}{n}\ =\ (-\frac{1}{2},-\frac{1}{2},-\frac{1}{2},-\frac{1}{2},+\frac{1}{2}), \\[2pt] 
\frac{\vec{a}_{23}}{n} &=& (+\frac{1}{2},-\frac{1}{2},-\frac{1}{2},+\frac{1}{2},+\frac{1}{2}), \hspace{20pt}
\frac{\vec{a}_{45}}{n}\ =\ (+\frac{1}{2},+\frac{1}{2},+\frac{1}{2},-\frac{1}{2},-\frac{1}{2}), \\[2pt] 
\frac{\vec{a}_{24}}{n} &=& (+\frac{1}{2},-\frac{1}{2},+\frac{1}{2},-\frac{1}{2},+\frac{1}{2}), \hspace{20pt}
\frac{\vec{a}_{35}}{n}\ =\ (+\frac{1}{2},+\frac{1}{2},-\frac{1}{2},+\frac{1}{2},-\frac{1}{2}), \\[2pt]
\frac{\vec{a}_{25}}{n} &=& (+\frac{1}{2},-\frac{1}{2},+\frac{1}{2},+\frac{1}{2},-\frac{1}{2}), \hspace{20pt}
\frac{\vec{a}_{34}}{n}\ =\ (+\frac{1}{2},+\frac{1}{2},-\frac{1}{2},-\frac{1}{2},+\frac{1}{2}). 
\end{eqnarray}

The commutation relations (\ref{comm_Q_x_5d}) imply commutation relations between
$\vec{x}$  and  $N=4\ D=5$ twisted Grassmann coordinates  
$(\theta,\theta_{\mu},\theta_{\mu\nu})$,
\begin{eqnarray}
[\vec{x}, \theta] \ =\ -\vec{a}\,  \theta,\ \ \ 
[\vec{x}, \theta_{\mu}] \ =\  -\vec{a}_{\mu}\,  \theta_{\mu},\ \ \ 
[\vec{x}, \theta_{\mu\nu}] \ =\  -\vec{a}_{\mu\nu}\,  \theta_{\mu\nu},\ \ \ 
 ({\rm \mu, \nu : no\ sum})\ \ \
\label{comm_theta_x_5d}
\end{eqnarray}
from which we may admit anti-commutation relations among the Grassmann coordinates,
\begin{eqnarray}
\{\theta,\theta_{\mu}\} &=& 
r^{(5)}_{a}\, g^{-1}_{+}(P_{\mu}) \ =\ 
-i\, r^{(5)}_{a}\, n\
e^{+i\vec{n}_{\mu}\cdot \vec{P}}
\label{N=4D=5tsusy1_lat_theta}\\[2pt]
\{\theta_{\rho\sigma},\theta_{\mu}\} &=& 
-r^{(5)}_{a}\, \delta_{\rho\sigma\mu\nu}\,
g^{-1}_{-}(P_{\nu}) \ =\ -i\, r^{(5)}_{a}\, n\, \delta_{\rho\sigma\mu\nu}
e^{-i\vec{n}_{\nu}\cdot \vec{P}} 
\label{N=4D=5tsusy2_lat_theta}\\[2pt]
\{\theta_{\rho\sigma},\theta_{\mu\nu}\} &=& 
+r^{(5)}_{a}\, \epsilon_{\rho\sigma\mu\nu\lambda}\,
g^{-1}_{+}(P_{\lambda}) \ =\
- i\, r^{(5)}_{a}\, n\, \epsilon_{\rho\sigma\mu\nu\lambda}
e^{+i\vec{n}_{\lambda}\cdot \vec{P}},
\label{N=4D=5tsusy3_lat_theta}
\end{eqnarray}
where $r^{(5)}_{a}$ representing a ratio parameter in the case of $N=4\ D=5$.
Correspondingly, anti-commutation relations among the derivatives of Grassmann coordinates may be introduced as,
\begin{eqnarray}
\bigl\{ \frac{\partial}{\partial\theta}, \frac{\partial}{\partial\theta_{\mu}}
 \bigr\} &=& t_{a}^{(5)} g_{+}(P_{\mu}), \\[5pt]
\bigl\{ \frac{\partial}{\partial\theta_{\rho\sigma}},
\frac{\partial}{\partial\theta_{\mu}}
 \bigr\} &=& - t_{a}^{(5)} \delta_{\rho\sigma\mu\nu}g_{-}(P_{\nu}), \\[5pt]
 \bigl\{ \frac{\partial}{\partial\theta_{\rho\sigma}},
\frac{\partial}{\partial\theta_{\mu\nu}}
 \bigr\} &=& + t_{a}^{(5)} \epsilon_{\rho\sigma\mu\nu\lambda}g_{+}(P_{\lambda}),
\end{eqnarray}
where $t_{a}^{(5)}$ representing another ratio parameter 
explained in the previous subsection, applied in the case of $N=4\ D=5$. 

\section{Applications to some particular cases}
\label{app_some_cases}

\indent

In general, multiplication properties of group elements with more than three parameters
do not fall into the category of the algebra  (\ref{XXY3})-(\ref{YYX3}).
However, in terms of the exponentially deformed twisted SUSY algebra, 
one may have chances to utilize the algebra  (\ref{XXY3})-(\ref{YYX3})
 and the closed expression of  BCH formula (\ref{BCH_closed1})-(\ref{BCH_closed3}) 
even for  group elements more then  three parameters.
In this section, we explore multiplication properties of
 group elements corresponding to the exponentially deformed twisted SUSY algebra
for some particular cases.

\subsection{$\alpha \neq 0,\ \beta = 0,\ \gamma=0$ in $D=N=2$}
\label{subsection_ps1}

\indent

First, we pick up a relatively simple case of $N=D=2$
with $r_{a}\neq0,\ r_{b}=r_{c}=0$, which
 means that we start from 
(\ref{N=2tsusy1_tr})-(\ref{N=2tsusy2_tr}) as $N=D=2$ deformed SUSY algebra
and (\ref{N=2tsusy1_tr_theta})-(\ref{N=2tsusy2_tr_theta}) as non(anti)commutativity
among $\theta 's$, $\{\theta_{A},\theta_{B}\} \neq 0$,
while keeping $\{\theta_{A}, \xi_{B}\} =
\{\xi_{A}, \xi_{B}\}  = 0$,
where $\theta_{A}$ denotes any of $(\theta, \theta_{\mu},\tilde{\theta})$,
while $\xi_{A}$ and $\xi_{B}$ denote any of $(\xi, \xi_{\mu},\tilde{\xi})$.

If we take X and Y as  
\begin{eqnarray}
 X &=& \xi Q + \xi_{\mu}Q_{\mu} + \tilde{\xi}\tilde{Q}, \ \ \ \
Y \ =\ \theta Q + \theta_{\mu}Q_{\mu} + \tilde{\theta}\tilde{Q}, 
\ \ \ (\mu = 1,2 {\rm :summed\ up})
\label{XY_2d}
\end{eqnarray}
then we have,
\begin{eqnarray}
[X,Y] &=& 
- \xi \theta_{\mu} g_{+}(P_{\mu})
- \xi_{\mu}\theta g_{+}(P_{\mu})
+ \xi_{\mu}\tilde{\theta}\epsilon_{\mu\nu} g_{-}(P_{\nu})
+ \tilde{\xi}\theta_{\mu}\epsilon_{\mu\nu} g_{-}(P_{\nu}), \label{comm_XY_2D}
\end{eqnarray}
from which we obtain,
\begin {eqnarray}
[X,[X,Y]] &=& 0, \label{XXY_2D_c1} \\[5pt]
[Y,[X,Y]] &=& \bigl[\theta Q,\, - \xi \theta_{\mu} g_{+}(P_{\mu})
+\tilde{\xi}\theta_{\mu}\epsilon_{\mu\nu} g_{-}(P_{\nu}) \bigr] \nonumber \\[5pt]
&& +\bigl[\theta_{\rho}Q_{\rho},\, - \xi_{\mu}\theta g_{+}(P_{\mu})
+ \xi_{\mu}\tilde{\theta}\epsilon_{\mu\nu} g_{-}(P_{\nu})\bigr] \nonumber  \\[5pt]
&& +\bigl[\tilde{\theta}\tilde{Q},\, - \xi \theta_{\mu} g_{+}(P_{\mu})
+\tilde{\xi}\theta_{\mu}\epsilon_{\mu\nu} g_{-}(P_{\nu})\bigr]   \\[5pt]
&=& +\xi Q r^{(2)}_{a} g^{-1}_{+} (P_{\mu})g_{+}(P_{\mu})
 - \tilde{\xi}Q r^{(2)}_{a} \epsilon_{\mu\nu}g^{-1}_{+}(P_{\mu})g_{-}(P_{\nu}) \nonumber\\[5pt]
&& +\xi_{\mu}Q_{\rho}r^{(2)}_{a} g^{-1}_{+} (P_{\rho})g_{+}(P_{\mu})
+ \xi_{\mu}Q_{\rho} r^{(2)}_{a}\epsilon_{\rho\sigma} g^{-1}_{-}(P_{\sigma})\epsilon_{\mu\nu}g_{-}(P_{\nu}) \nonumber \\[5pt]
&& -\tilde{\xi}\tilde{Q} r^{(2)}_{a}\epsilon_{\mu\lambda}g^{-1}_{-}(P_{\lambda})g_{+}(P_{\mu})
+\tilde{\xi}\tilde{Q} r^{(2)}_{a} \epsilon_{\mu\lambda} g^{-1}_{-}(P_{\lambda})
\epsilon_{\mu\nu}g_{-}(P_{\nu})  \label{YXY_2D_c2} \\[5pt]
&=& +2 r^{(2)}_{a} (\xi Q + \xi_{\mu} Q_{\mu} + \tilde{\xi} \tilde{Q}) 
\label{YXY_2D_c3} \\[5pt]
&=& +2 r^{(2)}_{a} X, \label{YXY_2D_c4}
\end{eqnarray}
where the second and fifth terms in (\ref{YXY_2D_c2}) vanish by themselves
due to 
$g^{-1}_{\pm}(P_{\mu}) = g_{\mp}(P_{\mu})$ and  the anti-symmetric propery of  
$\epsilon_{\mu\nu}$,  
while as for the fourth and sixth terms we have used
$\epsilon_{\rho\sigma}\epsilon_{\mu\nu} 
= \delta_{\rho\mu}\delta_{\sigma\nu}-\delta_{\rho\nu}\delta_{\sigma\mu}$.  
The above results (\ref{XXY_2D_c1}) and
(\ref{YXY_2D_c4}) indicate that the algebra regarding $X$ and $Y$ given in (\ref{XY_2d})
falls into the category of (\ref{XXY3})-(\ref{YYX3})
with $\alpha = -2 r^{(2)}_{a},\ \beta = \gamma =0$. 
More explicitly, we have for the BCH formula,
\begin{eqnarray}
e^{X}e^{Y}
&=& \exp \Bigl[ G(\alpha)X+Y+\frac{1}{2}[X,Y]\Bigr], \ \ \ \
\end{eqnarray}
where $X$ and $Y$ are given by (\ref{XY_2d}), while $G(\alpha)$ is given by
\begin{eqnarray}
G(\alpha) &=& 
\frac{\sqrt{\frac{1}{4}\alpha}\ \cosh \sqrt{\frac{1}{4}\alpha}}{\sinh \sqrt{\frac{1}{4}\alpha}}
\ =\ 
\frac{\sqrt{\frac{1}{2}r^{(2)}_{a}}\ 
}{\tan \sqrt{\frac{1}{2}r^{(2)}_{a}}}.
\end{eqnarray}
Thus, if we take the ratio parameter $r^{(2)}_{a}$ which satisfies,  
$\tan \sqrt{\frac{1}{2}r^{(2)}_{a}}=\sqrt{\frac{1}{2}r^{(2)}_{a}}$, then, the relation,
\begin{eqnarray}
e^{X}e^{Y}
&=& \exp \Bigl[ X+Y+\frac{1}{2}[X,Y]\Bigr] \ \ \ \ \label{X+Y+half} \label{BCH_exact_X+Y+half}
\end{eqnarray}
is exact.
The relation (\ref{X+Y+half}) gives rise to the supertranslations of the fermionic coordinates,
\begin{eqnarray}
\theta &\rightarrow& \theta +\xi,\ \ \ 
\theta_{\mu} \ \rightarrow\ \theta_{\mu} +\xi_{\mu}, \ \ \ 
\tilde{\theta} \ \rightarrow\ \tilde{\theta} +\tilde{\xi},  \label{var_ferm}
\end{eqnarray}
while providing a bosonic variation as
\begin{eqnarray}
\frac{1}{2}\Bigl[
- \xi \theta_{\mu} g_{+}(P_{\mu})
- \xi_{\mu}\theta g_{+}(P_{\mu})
+ \xi_{\mu}\tilde{\theta}\epsilon_{\mu\nu} g_{-}(P_{\nu})
+ \tilde{\xi}\theta_{\mu}\epsilon_{\mu\nu} g_{-}(P_{\nu})
\Bigr]. \ \ \label{var_boson}
\end{eqnarray}
The supertranslations of the fermionic coordinates (\ref{var_ferm}) and the 
expression (\ref{var_boson}) of the bosonic variation
may imply the differential forms of the supercharges $(Q,Q_{\mu},\tilde{Q})$
as below,
\begin{eqnarray}
Q &=&\frac{\partial}{\partial\theta}
-\frac{1}{2}\theta_{\mu} g_{+}(P_{\mu}), \label{Q_diff_N=D=2_1} \\
Q_{\mu} &=& \frac{\partial}{\partial\theta_{\mu}} 
-\frac{1}{2} \theta g_{+}(P_{\mu})
+\frac{1}{2} \tilde{\theta}\epsilon_{\mu\nu} g_{-}(P_{\nu}), \label{Q_diff_N=D=2_2}\\
\tilde{Q}&=& \frac{\partial}{\partial\tilde{\theta}}
+\frac{1}{2}\theta_{\mu}\epsilon_{\mu\nu} g_{-}(P_{\nu}). \label{Q_diff_N=D=2_3}
\end{eqnarray}
Noticing that the existence of non(anti)commutativities between $\theta_{A}$ and $\theta_{B}$, (\ref{N=2tsusy1_tr_theta})-(\ref{N=2tsusy2_tr_theta}),
as well as
non(anti)commutativities between $\frac{\partial}{\partial \theta_{A}}$ and  
$\frac{\partial}{\partial \theta_{B}}$, 
 (\ref{N=2tsusy1_tr_dtheta})-(\ref{N=2tsusy2_tr_dtheta}),
one can see that if $r_{a}^{(2)}$ and $t_{a}^{(2)}$ satisfy the condition, $t_{a}^{(2)}+\frac{1}{2}r_{a}^{(2)}=0$,
the differential forms of the supercharges (\ref{Q_diff_N=D=2_1})-(\ref{Q_diff_N=D=2_3}) satisfy
the corresponding anti-comutation relations,
\begin{eqnarray}
\{Q,Q_{\mu}\} &=& - g_{+}(P_{\mu}), \label{N=2tsusy1_tr_diff} \\[2pt]
\{\tilde{Q},Q_{\mu}\} &=& +\epsilon_{\mu\nu}g_{-}({P_{\nu}}).\label{N=2tsusy2_tr_diff}
\end{eqnarray}

Note here that in order to precisely investigate the supertranslation of bosonic coordinates, $x_{\mu}$,
we should consider how to parametrize the bosonic coordinates
in the super Lie group elements, which we haven't treated in this paper.  
Under the circumstances with the deformed SUSY algebra and the bosonic coordinates $x_{\mu}$ being regarded as symmetry generators of the deformed SUSY algebra, we may need further consideration of various possibilities of parametrizing the bosonic coordinates in the super Lie group elements under the entire algebraic structure.
Thus, we postpone this subject in the future investigation. 
Therefore, the expressions (\ref{Q_diff_N=D=2_1})-(\ref{Q_diff_N=D=2_3})
should be regarded as exemplary candidates of the differential forms of the supercharges,
or as the differential forms of the supercharges around the origin $x_{\mu}=0$.

Another interesting choice of the ratio parameter  $r^{(2)}_{a}$ is  to set,
$\tan \sqrt{\frac{1}{2}r^{(2)}_{a}}=\pm \infty$. In this choice, the relation,
\begin{eqnarray}
e^{X}e^{Y}
&=& \exp \Bigl[ Y+\frac{1}{2}[X,Y]\Bigr] \ \ \ \ \label{Y+half} 
\end{eqnarray}
is exact.
The relation (\ref{Y+half}) implies that the fermionic coordinates are unchanged,
\begin{eqnarray}
\theta &\rightarrow& \theta,\ \ \ 
\theta_{\mu} \ \rightarrow\ \theta_{\mu}, \ \ \ 
\tilde{\theta} \ \rightarrow\ \tilde{\theta},  \label{unchange_ferm}
\end{eqnarray}
while providing a bosonic variation as
\begin{eqnarray}
\frac{1}{2}\Bigl[
- \xi \theta_{\mu} g_{+}(P_{\mu})
- \xi_{\mu}\theta g_{+}(P_{\mu})
+ \xi_{\mu}\tilde{\theta}\epsilon_{\mu\nu} g_{-}(P_{\nu})
+ \tilde{\xi}\theta_{\mu}\epsilon_{\mu\nu} g_{-}(P_{\nu})
\Bigr], \ \ \label{var_boson2}
\end{eqnarray}
which implies that, in this choice of the ratio parameter, $r^{(2)}_{a}$, 
the  supercharges $(Q,Q_{\mu},\tilde{Q})$
may be expressed, without fermionic derivatives, as, 
\begin{eqnarray}
Q &\sim& 
\theta_{\mu} g_{+}(P_{\mu}), \label{Q_diff_N=D=2_21} \\[5pt]
Q_{\mu} &\sim& 
\theta g_{+}(P_{\mu})
-\tilde{\theta}\epsilon_{\mu\nu} g_{-}(P_{\nu}), \label{Q_diff_N=D=2_22}\\[5pt]
\tilde{Q}&\sim& 
\theta_{\mu}\epsilon_{\mu\nu} g_{-}(P_{\nu}). \label{Q_diff_N=D=2_23}
\end{eqnarray}
As noted above, since we haven't taken into account the 
 parametrization of the bosonic coordinates, $x_{\mu}$, the discussion should be taken as abstract. 
 Nevertheless, it is worthwhile to remind the redundant appearances of the translation operators,
 $g_{\pm}$,  mentioned at the end of subsection
\ref{deform_SUSY}. 
Here, the supercharges $(Q,Q_{\mu},\tilde{Q})$ are expressed without 
fermionic derivatives, which means that the anti-commutation relations between 
the supercharges may be obtained only using the  anti-commutation 
relations of the fermionic coordinates,
 (\ref{N=2tsusy1_tr_theta})-(\ref{N=2tsusy2_tr_theta}),
without consulting to the anti-commutation 
relations of the fermionic derivatives, 
 (\ref{N=2tsusy1_tr_dtheta})-(\ref{N=2tsusy2_tr_dtheta}).

\subsection{$\alpha \neq 0,\ \beta = 0,\ \gamma=0$ in $D=3\ N=4$,
$D=N=4$,  and  $D=5\ N=4$}
\label{subsection_ps2}

\indent

The same problem settings as above may be applied to  $D=3\ N=4$,
$D=N=4$,  and  $D=5\ N=4$. In the case of $D=3\ N=4$, we may define
\begin{eqnarray}
 X &=& \xi Q + \xi_{\mu}Q_{\mu} 
 + \overline{\xi}_{\mu}\overline{Q}_{\mu} 
 + \overline{\xi}\overline{Q}, \nonumber \\
Y & =& \theta Q + \theta_{\mu}Q_{\mu} 
+ \overline{\theta}_{\mu}\overline{Q}_{\mu}
+ \overline{\theta}\overline{Q}, 
\ \ \ (\mu = 1 \sim 3 {\rm :summed\ up}) 
\label{XY_3d}
\end{eqnarray}
from which we have,
\begin{eqnarray}
[X,Y] &=& -\xi\overline{\theta}_{\mu}g_{+}(P_{\mu})
-\xi_{\mu}\overline{\theta}g_{+}(P_{\mu})
+i\xi_{\mu}\overline{\theta}_{\nu}\epsilon_{\mu\nu\rho}\, g_{-}(P_{\rho}) 
\nonumber \\[5pt]
&& -\overline{\xi}_{\mu}\theta g_{+}(P_{\mu})
-i\overline{\xi}_{\mu}\theta_{\nu}\epsilon_{\mu\nu\rho}\, g_{-}(P_{\rho})
-\overline{\xi}\theta_{\mu}g_{+}(P_{\mu}), 
\end{eqnarray}
and
\begin{eqnarray}
[X,[X,Y]] &=&0, \label{XXY_3D} \\[5pt]
[Y,[X,Y]] &=& [\theta Q, -\xi\overline{\theta}_{\mu}g_{+}(P_{\mu})
+i\xi_{\mu}\overline{\theta}_{\nu}\epsilon_{\mu\nu\rho}\, g_{-}(P_{\rho})] 
\nonumber \\[5pt]
&& +[\theta_{\mu}Q_{\mu}, -\xi\overline{\theta}_{\nu}g_{}(P_{\nu})
-\xi_{\nu}\overline{\theta}g_{+}(P_{\nu})
+i\xi_{\alpha}\overline{\theta}_{\beta}\epsilon_{\alpha\beta\rho}\, g_{-}(P_{\rho})]
\nonumber \\[5pt]
&& +[\overline{\theta}_{\mu}\overline{Q}_{\mu}, -\overline{\xi}_{\nu}\theta g_{+}(P_{\nu})
-i\overline{\xi}_{\alpha}\theta_{\beta}\epsilon_{\alpha\beta\rho}\, g_{-}(P_{\rho})
-\overline{\xi}\theta_{\nu}g_{+}(P_{\nu}) ]
\nonumber  \\
&& + [\overline{\theta}\overline{Q}, -i\overline{\xi}_{\alpha}\theta_{\beta}\epsilon_{\alpha\beta\rho}\, g_{-}(P_{\rho})
-\overline{\xi}\theta_{\mu}g_{+}(P_{\mu}) ] \\[5pt]
&=& \xi Q r^{(3)}_{a}g^{-1}_{+}(P_{\mu})g_{+}(P_{\mu})
-i\xi_{\mu}Q r^{(3)}_{a}\epsilon_{\mu\nu\rho}\, g^{-1}_{+}(P_{\nu})g_{-}(P_{\rho}) 
\nonumber \\[5pt]
&& +i\xi Q_{\mu}r^{(3)}_{a}\epsilon_{\mu\nu\rho}\, g^{-1}_{-}(P_{\rho})g_{+}(P_{\nu})
+ \xi_{\nu}Q_{\mu}r^{(3)}_{a}g^{-1}_{+}(P_{\mu})g_{+}(P_{\nu})
\nonumber \\[5pt]
&& +\xi_{\alpha}Q_{\mu}r^{(3)}_{a}\epsilon_{\mu\beta\lambda}\epsilon_{\alpha\beta\rho}\,
g^{-1}_{-}(P_{\lambda})g_{-}P(\rho)
+\overline{\xi}_{\nu}\overline{Q}_{\mu}r^{(3)}_{a}g^{-1}_{+}(P_{\mu})g_{+}(P_{\nu})
\nonumber \\[5pt]
&& -\overline{\xi}_{\alpha}\overline{Q}_{\mu}r^{(3)}_{a}
\epsilon_{\beta\mu\lambda}
\epsilon_{\alpha\beta\rho}\, g^{-1}_{-} (P_{\lambda})g_{-}(P_{\rho})
+ i\overline{\xi}\overline{Q}_{\mu}r^{(3)}_{a}\epsilon_{\nu\mu\lambda}\,
g^{-1}_{-}(P_{\lambda})g_{+}(P_{\nu})
\nonumber \\[5pt]
&& + i\overline{\xi}_{\alpha}\overline{Q}r^{(3)}_{a}\epsilon_{\alpha\beta\rho}
g^{-1}_{+}(P_{\beta})g_{-}(P_{\rho})
+\overline{\xi}\overline{Q}r^{(3)}_{a}g^{-1}_{+}(P_{\mu})g_{+}(P_{\mu}) 
\label{YXY_3D_c2} \\[5pt]
&=& +3 r^{(3)}_{a}(\xi Q + \xi_{\mu}Q_{\mu}
+\overline{\xi}_{\mu}\overline{Q}_{\mu} + \overline{\xi}\overline{Q} )
\label{YXY_3D_c3}
\\[5pt]
&=& +3 r^{(3)}_{a} X, \label{YXY_3D_c4}
\end{eqnarray}
where the second, third, eighth and ninth terms in (\ref{YXY_3D_c2}) 
vanish by themselves due to 
$g^{-1}_{\pm}(P_{\mu}) = g_{\mp}(P_{\mu})$ and  the anti-symmetric propery of  
$\epsilon_{\mu\nu\rho}$,  
while as for the fifth and seventh terms we have used
$\epsilon_{\mu\beta\lambda}\epsilon_{\alpha\beta\rho} 
= \delta_{\mu\alpha}\delta_{\lambda\rho}-\delta_{\mu\rho}\delta_{\lambda\alpha}$.  
The above results (\ref{XXY_3D}) and (\ref{YXY_3D_c3})
indicate that the algebra regarding $X$ and $Y$ given in (\ref{XY_3d})
falls into the category of (\ref{XXY3})-(\ref{YYX3})
with $\alpha = -3 r^{(3)}_{a},\ \beta = \gamma =0$. 
More explicitly, we have for the BCH formula,
\begin{eqnarray}
e^{X}e^{Y}
&=& \exp \Bigl[ G(\alpha)X+Y+\frac{1}{2}[X,Y]\Bigr], \ \ \ \
\end{eqnarray}
where $X$ and $Y$ are given by (\ref{XY_3d}), while $G(\alpha)$ is given by
\begin{eqnarray}
G(\alpha) &=& 
\frac{\sqrt{\frac{1}{4}\alpha}\ \cosh \sqrt{\frac{1}{4}\alpha}}{\sinh \sqrt{\frac{1}{4}\alpha}}
\ =\ 
\frac{\sqrt{\frac{3}{4}r^{(3)}_{a}}\ 
}{\tan \sqrt{\frac{3}{4}r^{(3)}_{a}}}.
\end{eqnarray} 
Thus, if we take the ratio parameter $r^{(3)}_{a}$
which satisfies, $\tan\sqrt{\frac{3}{4}r^{(3)}_{a}}=\sqrt{\frac{3}{4}r^{(3)}_{a}}$,  
then, the relation,
\begin{eqnarray}
e^{X}e^{Y}
&=& \exp \Bigl[ X+Y+\frac{1}{2}[X,Y]\Bigr] \ \ \ \ \label{X+Y+half} \label{BCH_exact_X+Y+half}
\end{eqnarray}
is exact, while if we take,
$\tan \sqrt{\frac{3}{4}r^{(3)}_{a}}=\pm \infty$, the relation,
\begin{eqnarray}
e^{X}e^{Y}
&=& \exp \Bigl[ Y+\frac{1}{2}[X,Y]\Bigr] \ \ \ \ \label{Y+half} 
\end{eqnarray}
is exact.
Thus, regarding the differential forms of the supercharges, the similar arguments 
to the case of $D=N=2$ may be applied to the case of  
$N=4\ D=3$.

In a similar manner, in the  case of $D=N=4$, defining
\begin{eqnarray}
 X &=& \xi Q 
 + \xi_{\mu} Q_{\mu}
 +\frac{1}{2} \xi_{\mu\nu}Q_{\mu\nu} 
 + \tilde{\xi}_{\mu}\tilde{Q}_{\mu}
 + \tilde{\xi} \tilde{Q},
   \nonumber \\
Y & =& 
\theta Q 
 + \theta_{\mu} Q_{\mu}
 +\frac{1}{2} \theta_{\mu\nu}Q_{\mu\nu} 
 + \tilde{\theta}_{\mu}\tilde{Q}_{\mu}
 + \tilde{\theta} \tilde{Q}, 
\ \ \ (\mu,\nu = 1 \sim 4 {\rm :summed\ up}) 
\label{XY_4d}
\end{eqnarray}
one can repeat the same argument and obtain, after straightforward calculations,
\begin{eqnarray}
[X,[X,Y]] &=&0, \\[5pt]
[Y,[X,Y]] &=& + 4r^{(4)}_{a} X,
\end{eqnarray}
which means that  if we take 
$\tan\sqrt{r^{(4)}_{a}}=\sqrt{r^{(4)}_{a}}$,  
then, the relation,
\begin{eqnarray}
e^{X}e^{Y}
&=& \exp \Bigl[ X+Y+\frac{1}{2}[X,Y]\Bigr] \ \ \ \ \label{X+Y+half} \label{BCH_exact_X+Y+half}
\end{eqnarray}
is exact, while if we take,
$\tan \sqrt{r^{(4)}_{a}}=\pm \infty$, the relation,
\begin{eqnarray}
e^{X}e^{Y}
&=& \exp \Bigl[ Y+\frac{1}{2}[X,Y]\Bigr] \ \ \ \ \label{Y+half} 
\end{eqnarray}
is exact.
In the case of  $D=5\ N=4$, defining
\begin{eqnarray}
 X &=& \xi Q 
 + \xi_{\mu} Q_{\mu}
 +\frac{1}{2} \xi_{\mu\nu}Q_{\mu\nu} 
   \nonumber \\
Y & =& 
\theta Q 
 + \theta_{\mu} Q_{\mu}
 +\frac{1}{2} \theta_{\mu\nu}Q_{\mu\nu} 
\ \ \ (\mu,\nu = 1 \sim 5 {\rm :summed\ up}) 
\label{XY_5d}
\end{eqnarray}
we obtain, after straightforward calculations,
\begin{eqnarray}
[X,[X,Y]] &=&0, \\[5pt]
[Y,[X,Y]] &=& + 5r^{(5)}_{a} X.
\end{eqnarray}
which means that  if we take 
$\tan\sqrt{\frac{5}{4}r^{(5)}_{a}}=\sqrt{\frac{5}{4}r^{(5)}_{a}}$,  
then, the relation,
\begin{eqnarray}
e^{X}e^{Y}
&=& \exp \Bigl[ X+Y+\frac{1}{2}[X,Y]\Bigr] \ \ \ \ \label{X+Y+half} \label{BCH_exact_X+Y+half}
\end{eqnarray}
is exact, while if we take,
$\tan \sqrt{\frac{5}{4}r^{(5)}_{a}}=\pm \infty$, the relation,
\begin{eqnarray}
e^{X}e^{Y}
&=& \exp \Bigl[ Y+\frac{1}{2}[X,Y]\Bigr] \ \ \ \ \label{Y+half} 
\end{eqnarray}
is exact.

As seen above, even the simple non(anti)commutative problem setings already have 
rich non-linear structures which affect the superspace representasions of the supercharges.
Although we still need to clarify the parametrization of
the bosonic coordinates, the above simple calculations already have a considerable meaning to 
the formulation of lattice SUSY.
Namely, they imply that the Dirac-K\"ahler twisted $N=D=2$, $N=4\ D=3$, $N=D=4$
 and $N=4\ D=5$ SUSY algebra with the exponential deformation
 can be formulated exactly on  a lattice without any problem of infinite dimensionality
 of the superspace.
 The non-commutative problem setting taken in the above calculations may actually correspond to the 
 non-commutativite property embedded in the Dirac-K\"ahler twisted super Yang-Mills formulation 
on a lattice \cite{DKNS}.
 
For more general point of view, investigating other possible  problem settings should surely be 
interesting and necessary for the entire understanding of the formulation.
However, it is beyond the initial scope of this paper. 
Thus, instead of further looking into the ratio parameter space, 
we provide, in the next section, a geometrical aspect of the exponential deformation.

\section{Relation to link approach of twisted super Yang-Mills on a lattice}

\indent

So far, we have concentrated on a rigid superspace with non(anti)commutative
 Grassmann coordinates without introducing any interactions.
 However,  if we consider a gauge covariantization of the above framework,
 the geometrical meaning of the trivialization or exponential deformation of  
 the SUSY algebra becomes much clearer.

Let us consider a gauge covariantized version of the SUSY algebra  (\ref{SUSY_alg}),
\begin{eqnarray}
\{\D_{A}(x),\D_{B}(x)\} &=& \D_{AB}(x).
\label{gauged_alg}
\end{eqnarray}
In (\ref{gauged_alg}), $\D_{A}$ and $\D_{B}$ denote fermionic gauge covariant derivatice
defined by 
\begin{eqnarray}
\D_{A}(x) &=& D_{A} - i\Gamma_{A}(x,\theta_{A},\theta_{B}), \\[5pt]
\D_{B}(x) &=& D_{B} - i\Gamma_{B}(x,\theta_{A},\theta_{B}),
\end{eqnarray}
where $D_{A}$ and $D_{B}$ denote supercovariant derivatives corresponding to $Q_{A}$ and $Q_{B}$, respectively,
while  $\Gamma_{A}$ and $\Gamma_{B}$ representing
fermionic superconnections \cite{WZ_GSW_S}. 
In (\ref{gauged_alg}), $\D_{AB}$ denotes a bosonic gauge convariant derivatie defined by,
\begin{eqnarray}
\D_{AB}(x) &=& \partial_{AB} -i\Gamma_{AB}(x,\theta_{A},\theta_{B})
\end{eqnarray}
where $\partial_{AB}$ and
$\Gamma_{AB}$ representing a bosonic partial derivative and 
a bosonic superconnection, respectively. 
The fermionic and bosonic gauge covariant derivatives transform under 
 supergauge transformation in a local manner as,
\begin{eqnarray}
(\D_{A},\D_{B},\D_{AB}) &\rightarrow&
\mathcal{G}^{-1}(x) (\D_{A},\D_{B},\D_{AB})\,
\mathcal{G}(x)
\end{eqnarray}
where $\mathcal{G}$ denotes finite super gauge transformation which can be expressed in terms of arbitrary super field $\mathcal{K}$ as
$\mathcal{G} = e^{i\mathcal{K}}$.

An exponential deformation  of the 
gauged SUSY algebra (\ref{gauged_alg}) can be given by
\begin{eqnarray}
\{\D_{A}(x),\D_{B}(x)\} &=& -\frac{1}{n_{AB}}e^{-n_{AB}\D_{AB}(x)}, \label{exp_nabla1}
\ \ \  {\mathrm or} \\[5pt]
\{\D_{A}(x),\D_{B}(x)\} &=& +\frac{1}{n_{AB}}e^{+n_{AB}\D_{AB}(x)}. \label{exp_nabla2}
\end{eqnarray}
Note here that the right hand sides of (\ref{exp_nabla1}) and (\ref{exp_nabla2})
 transform locally under supergauge transformation,
since any k-th order of the bosonic
covariant derivative, $(\D_{AB})^{k}$, is still a locally covariant object,
\begin{eqnarray}
e^{\mp n_{AB}\D_{AB}(x)} &\rightarrow&
\mathcal{G}^{-1}(x)\, e^{\mp n_{AB}\D_{AB}(x)}\,
\mathcal{G}(x).
\end{eqnarray}
On the other hand, one may extract the contribution of shift,
$e^{\mp n_{AB}\partial_{AB}}$,
from the the right hand sides of (\ref{exp_nabla1}) and (\ref{exp_nabla2}) as
\begin{eqnarray}
e^{\mp n_{AB}\D_{AB}(x)} &=&
e^{\mp n_{AB}\D_{AB}(x)} e^{\pm n_{AB}\partial_{AB}} e^{\mp n_{AB}\partial_{AB}}
\label{derv_U_1}  \\[5pt]
&=& e^{\pm i n_{AB}\Gamma_{AB} -\frac{i}{2}n_{AB}^{2}[\partial_{AB} ,\Gamma_{AB}]
+\cdots}
e^{\mp n_{AB}\partial_{AB}} 
\label{derv_U_2} \\[5pt]
&=& (\mathcal{U}_{\pm})_{x,x\mp n_{AB}} \, e^{\mp n_{AB}\partial_{AB}}, 
\label{derv_U_3}
\end{eqnarray}
where, in (\ref{derv_U_2}), `$\cdots$' represents higer order terms including 
$\partial_{AB}$ and $\Gamma_{AB}$ which can be obtained by
Baker-Campbell-Hausdorff formula, while in  (\ref{derv_U_3}), we notationally introduced
gauge link variables as
\begin{eqnarray}
 (\mathcal{U}_{\pm})_{x,x\mp n_{AB}} &\equiv&
e^{\mp n_{AB}\D_{AB}(x)} e^{\pm n_{AB}\partial_{AB}} \\[5pt]
&=& e^{\pm i n_{AB}\Gamma_{AB} -\frac{i}{2}n_{AB}^{2}[\partial_{AB} ,\Gamma_{AB}]
+\cdots},
\end{eqnarray}
which have the following gauge covariant link properties
\begin{eqnarray}
 (\mathcal{U}_{\pm})_{x,x\mp n_{AB}} &\rightarrow&
 \mathcal{G}^{-1}(x)\,  (\mathcal{U}_{\pm})_{x,x\mp n_{AB}}\,
\mathcal{G}(x\mp n_{AB}).
\end{eqnarray}
In terms of the gauge link variables $\mathcal{U}_{\pm}$, the 
algebra (\ref{exp_nabla1}) and  (\ref{exp_nabla2}) can be expressed as
\begin{eqnarray}
\{\D_{A}(x),\D_{B}(x)\} &=& -\frac{1}{n_{AB}}
 (\mathcal{U}_{\pm})_{x,x- n_{AB}} e^{- n_{AB}\partial_{AB}} \\[5pt]
&=& -\frac{1}{n_{AB}}
e^{- n_{AB}\partial_{AB}} (\mathcal{U}_{\pm})_{x+n_{AB},x},
\label{exp_nabla_U1}
\ \ \  {\mathrm or} \\[5pt]
\{\D_{A}(x),\D_{B}(x)\} &=& +\frac{1}{n_{AB}}
 (\mathcal{U}_{\pm})_{x,x+ n_{AB}} e^{+ n_{AB}\partial_{AB}} \\[5pt] 
&=& +\frac{1}{n_{AB}}
e^{+ n_{AB}\partial_{AB}} (\mathcal{U}_{\pm})_{x-n_{AB},x}.
\label{exp_nabla_U2}
\end{eqnarray}
Here, the shift elements in the r.h.s can be consistently divided and rearranged
to each of $\D_{A}$ and $\D_{B}$ as
\begin{eqnarray}
e^{+ a_{A}\cdot \partial} \D_{A}(x+a_{B}) 
e^{+ a_{B}\cdot \partial} \D_{B}(x)  +
e^{+ a_{B}\cdot \partial} \D_{B}(x+a_{A}) 
e^{+ a_{A}\cdot \partial} \D_{A}(x)  
&=& -\frac{1}{n_{AB}}
(\mathcal{U}_{+})_{x+n_{AB},x}, \ \ \ \ \ \ \ \  \label{link_comm1}
\end{eqnarray}
in the case that $a_{A}+a_{B}=+n_{AB}$ is satisfied, or
\begin{eqnarray}
e^{+ a_{A}\cdot \partial} \D_{A}(x+a_{B}) 
e^{+ a_{B}\cdot \partial} \D_{B}(x)  +
e^{+ a_{B}\cdot \partial} \D_{B}(x+a_{A}) 
e^{+ a_{A}\cdot \partial} \D_{A}(x)  
&=& -\frac{1}{n_{AB}}
(\mathcal{U}_{-})_{x-n_{AB},x}, \ \ \ \ \ \ \ \ \label{link_comm2}
\end{eqnarray}
in the case that $a_{A}+a_{B}=-n_{AB}$ is satisfied.
Note that these manipulations correspond to the trivialization procedure
explained in the subsection \ref{deform_SUSY}
in the absence of the gauge fields.
Since the ``dressed'' fermionic covariant derivatives in the left hand sides
of (\ref{link_comm1})  and (\ref{link_comm2})
have the following link gauge covariant properties,
\begin{eqnarray}
e^{+ a_{A}\cdot \partial} \D_{A}(x)   &\rightarrow&
 \mathcal{G}^{-1}(x+a_{A})\,  e^{+ a_{A}\cdot \partial} \D_{A}(x) \, 
  \mathcal{G}^{-1}(x), \\[5pt]
 e^{+ a_{B}\cdot \partial} \D_{A}(x)   &\rightarrow&
 \mathcal{G}^{-1}(x+a_{B})\,  e^{+ a_{B}\cdot \partial} \D_{B}(x) \, 
  \mathcal{G}^{-1}(x)
\end{eqnarray}
it is appropriate to introduce the following fermionic link notations,
\begin{eqnarray}
(\D_{A})_{x+a_{A},x} &\equiv& e^{+ a_{A}\cdot \partial} \D_{A}(x),
\ \ \ \ \  
(\D_{B})_{x+a_{B},x} \ \equiv\  e^{+ a_{B}\cdot \partial} \D_{B}(x).
\end{eqnarray}
with which the gauged SUSY algebra (\ref{link_comm1}) and (\ref{link_comm1})
can be re-expressed in terms of a notion of  ``link-(anti)commutators''
of the fermionic link variables as, 
\begin{eqnarray}
 \{\D_{A},\D_{B}\}_{x+a_{A}+a_{B},x}
 &\equiv&
(\D_{A})_{x+a_{A}+a_{B},x+a_{B}} (\D_{B})_{x+a_{B},x}
+ (\D_{B})_{x+a_{B}+a_{A},x+a_{A}} (\D_{A})_{x+a_{A},x}  \nonumber \\[5pt]
&=& 
 -\frac{1}{n_{AB}}
(\mathcal{U}_{+})_{x+n_{AB},x},  \label{link_comm11}
\\[5pt]
 \{\D_{A},\D_{B}\}_{x+a_{A}+a_{B},x}
 &=&
(\D_{A})_{x+a_{A}+a_{B},x+a_{B}} (\D_{B})_{x+a_{B},x}
+ (\D_{B})_{x+a_{B}+a_{A},x+a_{A}} (\D_{A})_{x+a_{A},x}  \nonumber \\[5pt]
&=& 
 -\frac{1}{n_{AB}}
(\mathcal{U}_{-})_{x-n_{AB},x}, \label{link_comm12}
\end{eqnarray}
respectively. 
Fig. 1 and Fig. 2 show the geometric configurations of the 
fermionic and bosonic gauge link variables in the 
gauged SUSY algebra (\ref{link_comm11}) and  (\ref{link_comm12}), respectively.

These  ``link-(anti)commutators'' have been the essential algebraic
framework in formulating Dirac-K\"ahler twisted super Yang-Mills
on a lattice in  various dimensions \cite{DKKN2, DKKN3,DKNS}.
For example, $N=4\ D=5$ twisted super Yang-Mills constraints on a lattice
are given by
\begin{eqnarray}
\{\D,\D_{\mu}\}_{x+a+a_{\mu},x} &=& +i(\U_{+\mu})_{x+n_{\mu},x} \label{c1_lat}\\[2pt]
\{\D_{\rho\sigma},\D_{\mu}\}_{x+a_{\rho\sigma}+a_{\mu},x}
 &=& +i\delta_{\rho\sigma\mu\nu}(\U_{-\nu})_{x-n_{\nu},x}\label{c2_lat} \\[2pt]
\{\D_{\rho\sigma},\D_{\mu\nu}\}_{x+a_{\mu\nu}+a_{\rho\sigma}} &=& + i\epsilon_{\rho\sigma\mu\nu\lambda}(\U_{+\lambda})_{x+n_{\lambda},x}.
\label{c3_lat}
\end{eqnarray}
where the bosonic gauge link variables can be represented as
\begin{eqnarray}
 (\U_{\pm\mu})_{x\pm n_{\mu},x} \ = \ 
 (e^{\pm i (A_{\mu}\pm i\phi^{(\mu)}+\cdots)})_{x\pm n_{\mu},x}.
\end{eqnarray}
where $A_{\mu}$ and $\phi^{(\mu)}$ denoting the bosonic superfields
whose loewst components include contributions of 
five dimensional gauge fields and five components of scalar fields
 in the $N=4\ D=5$ SYM multiplet.
  
After seeing the above derivation, it is natural to understand that
the trivialized algebra (\ref{triv_Q_alg}) corresponds to 
the one where the gauge felds are tuned off, thus,
the left hand side of  (\ref{triv_Q_alg})  should be treated as a link anti-commutator.

\begin{figure}
\begin{center}
\begin{minipage}{65mm}
\begin{center}
\includegraphics[width=50mm]{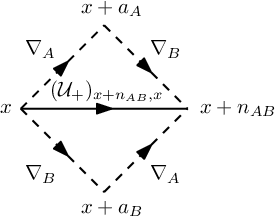}
\caption{Geometric configuration of the 
fermionic and bosonic gauge link variables in the 
gauged SUSY algebra (\ref{link_comm11}) }
\label{s_Delta1}
\end{center}
\end{minipage}
\hspace{20pt}
\begin{minipage}{65mm}
\begin{center}
\includegraphics[width=50mm]{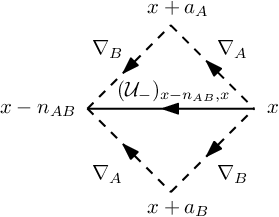}
\caption{Geometric configuration of the 
fermionic and bosonic gauge link variables in the 
gauged SUSY algebra (\ref{link_comm12})}
\label{s_Delta2}
\end{center}
\end{minipage}
\end{center}
\end{figure}

\section{Summary and Discussions}

\indent

Starting from a very elementary calculation,
we have seen that
the non(anti)commutativity
between the Grassmann coordinates in a vector sector
 in general brings about
the infinite dimensionality of the algebra in the superspace.
In order to circumvent such an infinite dimensionality, we have introduced a 
notion of exponential deformation or trivialization of the algebra,
and seen that several specific algebra satisfy the deformation conditions.
In particular, by means of the Baker-Campbell-Haussdorff closed formula,
we have explicitly seen that the exponentially deformed $N=D=2$, $N=4\ D=3$, $N=D=4$ and $N=4\ D=5$ Dirac-K\"ahler twisted SUSY algebra
can circumvent the infinite dimensionality and provides a feasible supertranslation property in the superspace.
Although there are still a lot of aspects to be clarified, such as 
a vast parameter ($\alpha$, $\beta$, $\gamma$) space of the possible problem settings of the non(anti)commutativity,
and a parametrization of the bosonic coordinates, etc., 
the present framework provides a novel and practical methodology to the
study of the non(anti)commutative superspace.
There are many things to be addressed.
Several comments and discussions are in order.

\subsection{Providing a basic framework for Lattice SUSY}

\indent 

As already mentioned, the exponential deformation provides a basic framework to  realize the lattice SUSY. 
A discretization of spacetime through the exponential deformation
may provides a feasible methodology to  circumvent the infinite dimensionality resulted from the non(anti)commutativity among the Grassmann coordinates.
It should be stressed that in order to consistently carry out the exponential deformation, the algebra should satisfy the deformation condition which is actually what we have been calling lattice Liebniz rule condition.
According to our study, in order to satisfy the  lattice Liebniz rule condition, 
the SUSY algebra should be twisted in a very specific way, which is Dirac-K\"ahler twisting (Marcus B-type twisting or Geometric Langlands twisting).
For example, a comparison between A-type twist and B-type twist 
from a perspective of lattice super Yang-MIlls
implementation is given in \cite{Nagata1}.

It is important to point out that, as we have seen in the section \ref{DK_alg_various}, 
after Dirac-K\"ahler twisting, the supercharges behave as a scalar, vector, tensor, ... 
each of which has one-to-one correspondence to the simplicial element in the corresponding dimension. This implies that non(anti)commutativity among the Grassmann coordinates may naturally leads to a simplicial discretization of  spacetime. 
As already mentioned, the exemplary non(anti)commutative setting studied in the section 
\ref{app_some_cases} may correspond to the non-commutative setting in the gauge covariant link formulations of twisted super Yang-Mills on a lattice \cite{DKNS}. 
Therefore, the non(anti)commutative superspace examined in this article may serve as an
underlying superspace framework for lattice SUSY.

It is also interesting to ask whether the ratio parameters $r_{a}$, $r_{b}$ and $r_{c}$
have any physical meanings or physical interpretations
in the lattice implementation of super Yang-Mills.
We also note that,
from a viewpoint of regularization, there have been a large amount of lattice SUSY studies,
some of which are for example 
\cite{Catterall_Giedt_Toga, Schaich, Kato_Sakamoto_So, Kadoh, Soler, Sugino, Matsuura_Sugino,Unsal},
as well as a large amount of non-lattice  studies,
some of which are for example 
\cite{Ohta_Matsuura, Hirasawa_etal, Ito_Nakajima_Sasaki, Halimeh}.
It is interesting to consider how and what the present formulation explored in this paper 
may affect to those related topics.

\subsection{Possibility for applying to the chiral or anti-chiral sector}

\indent

As already explained, the entire non-commutative problem setting examined in this paper
belongs to the vector sector in the sense that we deal with $\{\theta_{A},\theta_{B}\}=a_{AB}$
associating to the corresponding SUSY algebra $\{Q_{A},Q_{B}\}=P_{AB}$
From the context of string theory, it is important to ask whether the present formulation
can be applied to the chiral sector $\{\theta^{\alpha},\theta^{\beta}\}=C^{\alpha\beta}$
or anti-chiral sector $\{\theta^{\dot{\alpha}},\theta^{\dot{\beta}}\}=C^{\dot{\alpha}\dot{\beta}}$.
Recognizing that the corresponding graviphoton field $F^{\alpha\beta}$ or $F^{\dot{\alpha}\dot{\beta}}$
may be regarded as a second-rank tensor,
the problem may include asking how to decompose a bosonic face element
consistently with the corresponding SUSY algebra.

It is also worthwhile to mention that the exponential deformation
examined in this paper 
implies a scenario where the non-commutativity among the Grassmann parameters 
and the one among the supercharges are inverse each other.
More specifically, it may imply a scenario that 
for any reason, if  a non(anti)commutativity 
between the Grassmann coordinates arises with some amount $G(a_{AB})$, namely,
 $\{\theta_{A},\theta_{B}\} = G(a_{AB})$,
then such a non(anti)commutativity may induce a corresponding
SUSY algebra 
which can be regarded as a fermionic decomposition
 of the inverse of the non(anti)commutativity $G(n_{AB})$, namely,
 $\{Q_{A},Q_{B}\}=G^{-1}(a_{AB})$.

%

\subsection{Baker-Campbell-Haussdorff closed formulae}

\indent

As shown in the Appendix, we have explicitly derived the Baker-Campbell-Haussdorff closed formulae and utilized them in the analysis of the non(anti)comutative superspace.
The BCH formula itself has been attracting a lot of interests of its own, and
playing a fundamental role in Lie group analyses or in other subjects.   
There have been several studies reporting exact  BCH formulae \cite{Van-Brunt1, Matone1, Moodie, Zadra}  
and their various applications to theoretical and experimental physics
 \cite{Matone_Pasti, Das, Liska, Caputa, Kataev, Eckhardt, Weiss}.
To the author's understanding, the general BCH closed form derived in this paper may correspond to the one which was reported in \cite{Zadra} in a different context, although  the underlying commutation relations were not explicitly presented
in \cite{Zadra}.
We also note that a specific case of the 
general BCH closed form derived in this paper corresponds to a formula presented in  \cite{Matone1}.
In this paper, 
as shown in the Appendix, the derivation is based on 
a sort of diagonalized tensor algebra 
where  each of the elements  is consisted by 
a product of diagonalized non(anti)commutative Grassmann coordinates and supercharges, $\theta^{\pm} \otimes Q^{\pm}$.
Therefore, it is natural to expect that  we may have a chance to obtain
more general BCH closed forms, starting from a higer rank tensor algebra.

\subsection{Investigating entire algebraic structure including rotational symmetries}

\indent

In this paper, we haven't discussed about how the spacetime roational symmetry $J$,
the internal rotational symmetry $R$, and the twisted Lorentz symmetry 
$J' = J+R$ are affected
by the exponential deformation, or in other words, by the discretization of spacetime.
On the other hand,
we have seen  that, in association with the exponential deformation, the scale
operator $\bf{D}$ superficially splits to the coordinates $x_{\mu}$ as symmetry generators, and the 
 Weyl - `tHooft algebra between $P_{AB}$ and $x$  naturally emerges.
These aspects imply that  a rich algebraic and group structure may exist behind this formulation. 
We observe that above-mentioned 
rotational symmetries should be examined along this line.

\subsection{Relation to theta function, and locaity v.s. non-locality}

\indent

As a bit more mathematical point of view,
in \cite{Kapranov}, with reference to \cite{Sato1},
a role of supergroup like element $e^{Q}$ with $Q^{2}=\frac{\partial}{\partial t}$ was 
discussed from an aspect of symmetry generators of 
periodicity properties (or symmetric properties) of the theta function,
\begin{eqnarray}
\theta(t,z) = \sum_{n\in Z}
e^{n^{2}t +  nz},
\end{eqnarray}
and the locality of the element $e^{Q}$ was also discussed. It is interesting to ask how the relations or definitions are modified or extended if we take into account the non(anti)commutative Grassmann parameter $\theta_{A}$'s or
a diagonal version of them $\theta^{\pm}$ introduced in the Appendix.
It is also interesting to see how the discussion of locality and non-locality in the above context may be affected by the exponential deformation of the SUSY algebra.

\subsection{Relation to Bernoulli generating function and polynomials}

\indent

At last, regarding the Baker-Campbell-Haussdorff closed forms, it is worthwhile to
mention their relation to Bernoulli generating function and polynomials.
The factor $F(\alpha,\beta,\gamma)$ of (\ref{BCH_closed2})
can be expressed as
\begin{eqnarray}
F(\alpha, \beta,\gamma) 
&=&
\frac
{\sinh\sqrt{\frac{1}{4}\alpha}\ \sinh\sqrt{\frac{1}{4}\gamma} }
{\sqrt{\frac{1}{4}\alpha} \sqrt{\frac{1}{4}\gamma} } 
\frac{\log (s + \sqrt{s^{2}-1}) }
{\sqrt{s^{2}-1}} 
\label{factor_F1}
\end{eqnarray}
where $s$ is given by
\begin{eqnarray}
s &=&
\cosh \sqrt{\frac{1}{4}\alpha}\, \cosh \sqrt{\frac{1}{4}\gamma} 
- \frac{\beta}{\sqrt{\alpha\gamma}}\, \sinh \sqrt{\frac{1}{4}\alpha}\, \sinh \sqrt{\frac{1}{4}\gamma} .
\end{eqnarray}
Here, the log factor in (\ref{factor_F1})
can be re-expressed as,
 in terms of $u=\cosh^{-1}s$,
\begin{eqnarray}
\frac{\log (s + \sqrt{s^{2}-1}) }
{\sqrt{s^{2}-1}} 
&=&  \frac{2u}{e^{u}-e^{-u}} 
\ =\ \frac{2u e^{u}}{e^{2u}-1}
\end{eqnarray}
which should be compared with the Bernoulli generating function,
\begin{eqnarray}
\frac{t e^{xt}}{e^{t}-1} &=& \sum_{n=0}^{\infty} B_{n}(x)\frac{t^{n}}{n!}.
\end{eqnarray}
where $B_{n}(x)$ represents Bernoulli polynomials. 
Identifying $t=2u$ and $x=\frac{1}{2}$, we obtain a Bernoulli polynomial expression of
$F(\alpha,\beta,\gamma)$ as
\begin{eqnarray}
F(\alpha,\beta,\gamma)
&=&  
\frac
{\sh\sqrt{\frac{1}{4}\alpha}\ \sh\sqrt{\frac{1}{4}\gamma} }
{\sqrt{\frac{1}{4}\alpha} \sqrt{\frac{1}{4}\gamma} } 
\sum_{n=0}^{\infty} B_{n}(\frac{1}{2})\frac{2^{n}u^{n}}{n!}.
\end{eqnarray}

\section*{Acknowledgments}

The author would like to thank 
 N. Kawamoto and J. Saito 
for the discussions and comments in particular at the early stage of this work.

\appendix

\section{Explicit derivation of Baker-Campbell-Hausdorff closed forms}
\renewcommand{\theequation}{A.\arabic{equation}}
\setcounter{equation}{0}

\indent

In this appendix, we explicitly derive the Baker-Campbell-Hausdorff closed forms
inncluding 
(\ref{BCH_closed1})-(\ref{BCH_closed3}) 
associated with (\ref{XXY3})-(\ref{YYX3}),
and (\ref{XY_BCH1})-(\ref{XY_BCH2}) associated with   (\ref{XXY1})-(\ref{YYX1}).
Let us begin with the non(anti)commutative Grassmann parameters $\theta_{A,B}$ and $\xi_{A,B}$ which satisfy
\begin{eqnarray}
\{\theta_{A},\theta_{B}\} &=& a_{AB}, \ \ \ \ \theta_{A}^2 \ =\ \theta_{B}^2 \ =\ 0, \label{NAC_app}\\[5pt]
\{\xi_{A},\xi_{B}\} &=& c_{AB}, \ \ \ \ \xi_{A}^2 \ =\ \xi_{B}^2 \ =\ 0,\\[5pt]
\{\xi_{A},\theta_{B}\} &=& \{\xi_{B},\theta_{A}\} =  b_{AB}, \ \ \ \ \{\xi_{A},\theta_{A}\} = \{\xi_{B},\theta_{B}\} = 0 \ \ \ \ (A, B:\rm{no\ sum}),
\end{eqnarray}
and the supercharges $Q_{A,B}$ which satisfy
\begin{eqnarray}
\{Q_{A},Q_{B}\} &=& P_{AB}, \ \ \ Q_{A}^2 \ =\ Q_{B}^2 \ =\ 0, \label{SUSY_alg_app} \\[5pt]
\{\theta_{A}, Q_{A}\} &=& \{\theta_{B}, Q_{B}\} \ =\ \{\theta_{A}, Q_{B}\} \ =\ \{\theta_{B}, Q_{A}\} \ =\ 0,
\ \ (A, B:\rm{no\ sum}), \\[5pt]
\{\xi_{A}, Q_{A}\} &=& \{\xi_{B}, Q_{B}\} \ =\ \{\xi_{A}, Q_{B}\} \ =\ \{\xi_{B}, Q_{A}\} \ =\ 0.
\ \ (A, B:\rm{no\ sum}).
\end{eqnarray}
We then introduce the following diagonal basis $\theta^{\pm}$, $\xi^{\pm}$ and $Q^{\pm}$,
\begin{eqnarray}
\theta^{+} &\equiv& \frac{1}{\sqrt{2}} (\theta_{A}+\theta_{B}),\ \ \ \ 
\theta^{-} \ \equiv\ \frac{1}{\sqrt{2}i} (\theta_{A}-\theta_{B}),\\[5pt]
\xi^{+} &\equiv& \frac{1}{\sqrt{2}} (\xi_{A}+\xi_{B}),\ \ \ \ 
\xi^{-} \ \equiv\ \frac{1}{\sqrt{2}i} (\xi_{A}-\xi_{B}),\\[5pt]
Q^{+} &\equiv& \frac{1}{\sqrt{2}} (Q_{A}+Q_{B}), \ \ \ \ 
Q^{-} \ \equiv\ \frac{1}{\sqrt{2}i} (Q_{A}-Q_{B}),
\end{eqnarray}
which satisfy
\begin{eqnarray}
\{ \theta^{+}, \theta^{-}\} &=& 0,\ \ \ \ 
(\theta^{+})^2 = (\theta^{-})^2 \ =\ \frac{1}{2} a_{AB},\\[2pt]
\{ \xi^{+}, \xi^{-}\} &=& 0,\ \ \ \ 
(\xi^{+})^2 = (\xi^{-})^2 \ =\ \frac{1}{2} c_{AB}, \\[5pt] 
\{ \xi^{+}, \theta^{+}\} &=& \{ \xi^{-}, \theta^{-}\} \ =\ b_{AB}, \ \ \ \ 
\{ \xi^{+},\theta^{-} \} \ =\ \{\xi^{-},\theta^{+}\} \ =\ 0,
\\[5pt]
\{ Q^{+}, Q^{-}\} &=& 0,\ \ \ \ \
(Q^{+})^2 = (Q^{-})^2 \ =\ \frac{1}{2} P_{AB}.
\end{eqnarray}
In terms of the diagonal basis, if we define,
\begin{eqnarray}
X \equiv \xi^{+}Q^{+}, \ \ \ \ Y \ \equiv\ \theta^{+}Q^{+}, \label{AandY_app}
\end{eqnarray}
one can see that the above elements $X$ and $Y$ satisfy the 
same commutation relations as Eqs. (\ref{XXY3}) and (\ref{YYX3}),
\begin{eqnarray}
[X,[X,Y]] &=& \gamma Y + \beta X, \ \ \rm{and} \label{XXY3_app} \\[5pt]
[Y,[Y,X]] &=& \alpha X + \beta Y, \label{YYX3_app} 
\end{eqnarray}
where
\begin{eqnarray}
\alpha &\equiv& - a_{AB}P_{AB}, \ \ \ \
\beta \ \equiv\ + b_{AB}P_{AB}, \ \ \ \ 
\gamma \ \equiv\ - c_{AB}P_{AB}.
\end{eqnarray}
We then try to find solutions 
$\lambda$, $\sigma'$ and $\tau'$ which satisfy
the following closed relaion,
\begin{eqnarray}
e^{\sigma X} e^{\tau Y} &=& \exp \Bigl( {\sigma'X+\tau'Y +\frac{1}{2}\lambda [X,Y]} \Bigr) \label{BCH_app}
\end{eqnarray}
for arbitrary $\sigma$ and $\tau$.
As for the l.h.s. of  (\ref{BCH_app}), we have
\begin{eqnarray}
e^{\sigma X} e^{\tau Y} &=& \Bigl( \cosh{\sigma \sqrt{\frac{1}{4}\gamma}} + \frac{X}{\sqrt{\frac{1}{4}\gamma}} \sinh{\sigma \sqrt{\frac{1}{4}\gamma}} \Bigr)
\Bigl( \cosh{\tau \sqrt{\frac{1}{4}\alpha}} + \frac{Y}{\sqrt{\frac{1}{4}\alpha}} \sinh{\tau \sqrt{\frac{1}{4}\alpha}}\Bigr), \label{BCHL_app}
\end{eqnarray}
where we utilized the property, $X^{2}=\frac{1}{4}\gamma$, and $Y^{2}=\frac{1}{4}\alpha$, obtained from
(\ref{AandY_app}).
As for the r.h.s. of  (\ref{BCH_app}), after some calculations, we have for the argumant of the exponential,
\begin{eqnarray}
(\sigma' X + \tau' Y +\frac{1}{2}\lambda[X,Y])^{2} &=&
\frac{1}{4} \sigma'^{2}\gamma + \frac{1}{4}\tau'^{2}\alpha
-\frac{1}{2}\sigma'\tau' \beta
+\frac{1}{16}\lambda^2 (-\alpha\gamma + \beta^{2}) 
\ \equiv\ d, \label{def_d_app}
\end{eqnarray}
where we utilized several properties obtained from (\ref{AandY_app}), and introduced a tentative parameter $d$  for notational simplicity.
Note here that, as shown above, 
in terms of the explicit representations of  (\ref{AandY_app}),
the square of the argument does not depend on
$X$ or $Y$, which makes it possible to express the r.h.s. of  (\ref{BCH_app}) as a closed form as follows,
\begin{eqnarray}
\exp \Bigl( {\sigma'X+\tau'Y +\frac{1}{2}\lambda [X,Y]} \Bigr) 
&=&
\cosh  \sqrt{d} + \frac{1}{\sqrt{d}}\Bigl(\sigma'X+\tau'Y +\frac{1}{2}\lambda [X,Y]\Bigr) 
\sinh  \sqrt{d}.
\ \ \ \label{BCHR_app}
\end{eqnarray}
Noticing that $XY$ and $[X,Y]$ are given as,
\begin{eqnarray}
XY &=& -\frac{1}{2}\xi^{+}\theta^{+}P_{AB}, \ \ \ \ [X,Y] \ =\ -\xi^{+}\theta^{+}P_{AB}+\frac{1}{2}\beta,
\end{eqnarray}
and equating  the right hand sides of (\ref{BCHL_app}) and (\ref{BCHR_app}), we have,
\begin{eqnarray}
\cosh \sigma \sqrt{\frac{1}{4}\gamma} \cosh \tau \sqrt{\frac{1}{4}\alpha}
&=& \cosh  \sqrt{d} + \frac{1}{4\sqrt{d}} \lambda \beta 
\sinh  \sqrt{d}, \label{BCHrel1_app}\\[5pt]
\frac{1}{\sqrt{\frac{1}{4}\alpha}}\cosh \sigma\sqrt{\frac{1}{4}\gamma}
\sinh \tau\sqrt{\frac{1}{4}\alpha}
&=& \frac{1}{\sqrt{d}}\tau' \sinh\sqrt{d}, \label{BCHrel2_app}\\
\frac{1}{\sqrt{\frac{1}{4}\gamma}}\sinh \sigma\sqrt{\frac{1}{4}\gamma}
\cosh \tau\sqrt{\frac{1}{4}\alpha}
&=& \frac{1}{\sqrt{d}}\sigma' \sinh\sqrt{d}, \label{BCHrel3_app} \\
\frac{1}{\sqrt{\frac{1}{4}\gamma}}\frac{1}{\sqrt{\frac{1}{4}\alpha}}\sinh \sigma\sqrt{\frac{1}{4}\gamma}
\sinh \tau\sqrt{\frac{1}{4}\alpha}
&=& \frac{1}{\sqrt{d}}\lambda \sinh\sqrt{d}. \label{BCHrel4_app}
\end{eqnarray}
From (\ref{BCHrel2_app}) and (\ref{BCHrel3_app}), we obtain for a ratio of $\tau'$ to $\sigma'$ as,
\begin{eqnarray}
\frac{\tau'}{\sigma'} 
&=& 
\frac{\sqrt{\frac{1}{4}\gamma}
\cosh \sigma\sqrt{\frac{1}{4}\gamma}\sinh \tau\sqrt{\frac{1}{4}\alpha}}
{\sqrt{\frac{1}{4}\alpha}\sinh \sigma\sqrt{\frac{1}{4}\gamma}
\cosh \tau\sqrt{\frac{1}{4}\alpha}}, \label{BCHrel5_app}
\end{eqnarray}
while from (\ref{BCHrel2_app}) and (\ref{BCHrel4_app}) as well as  (\ref{BCHrel3_app}) and (\ref{BCHrel4_app}), we obtain
\begin{eqnarray}
\frac{\tau'}{\lambda} &=& \frac{\sqrt{\frac{1}{4}\gamma}
\cosh \sigma\sqrt{\frac{1}{4}\gamma}}{\sinh \sigma\sqrt{\frac{1}{4}\gamma}}, \ \ \ \ 
\frac{\sigma'}{\lambda} \ =\ \frac{\sqrt{\frac{1}{4}\alpha}
\cosh \tau\sqrt{\frac{1}{4}\alpha}}{\sinh \tau\sqrt{\frac{1}{4}\alpha}}. \label{BCHrel6_app}
\end{eqnarray}
As for (\ref{BCHrel1_app}), substituting (\ref{BCHrel2_app}) or (\ref{BCHrel3_app}) into the second term of the
r.h.s. of (\ref{BCHrel1_app}) and using (\ref{BCHrel6_app}), we obtain
\begin{eqnarray}
\sqrt{d}&=& 
\cosh^{-1}\Biggl(
\cosh \sigma\sqrt{\frac{1}{4}\gamma}\cosh \tau\sqrt{\frac{1}{4}\alpha}
- \frac{\beta}{\sqrt{\alpha\gamma}}
\sinh \sigma\sqrt{\frac{1}{4}\gamma}\sinh \tau\sqrt{\frac{1}{4}\alpha}
\Biggr).
\end{eqnarray}
Remebering that $d$ is given by (\ref{def_d_app}) and utilizing the relations (\ref{BCHrel6_app})
and $\cosh^{-1}x = \log(x+\sqrt{x^2 -1})$, we have
\begin{eqnarray}
&& \lambda = \nonumber  \\
&& \frac{\log \Bigl( 
\ch\sigma \sqrt{\frac{1}{4}\gamma}\, \ch\tau \sqrt{\frac{1}{4}\alpha} 
- \frac{\beta}{\sqrt{\alpha\gamma}}\, \sh\sigma \sqrt{\frac{1}{4}\gamma}\, \sh\tau \sqrt{\frac{1}{4}\alpha} + \sqrt{\bigl(\ch\sigma \sqrt{\frac{1}{4}\gamma}\, \ch\tau \sqrt{\frac{1}{4}\alpha} 
- \frac{\beta}{\sqrt{\alpha\gamma}}\, \sh\sigma \sqrt{\frac{1}{4}\gamma}\, \sh\tau \sqrt{\frac{1}{4}\alpha} 
\bigr)^2 -1}\Bigr)}
{\sqrt{\frac{1}{16}\alpha\gamma
\Bigl(
\frac{\ch^2 \tau\sqrt{\frac{1}{4}\alpha}}{\sh^2 \tau\sqrt{\frac{1}{4}\alpha}}
+\frac{\ch^2 \sigma\sqrt{\frac{1}{4}\gamma}}{\sh^2 \sigma\sqrt{\frac{1}{4}\gamma}}
\Bigr)
-\frac{1}{8}\beta\sqrt{\alpha\gamma}\,
\frac{\ch \sigma\sqrt{\frac{1}{4}\gamma}\, \ch \tau\sqrt{\frac{1}{4}\alpha}}
{\sh \sigma\sqrt{\frac{1}{4}\gamma}\, \sh \tau\sqrt{\frac{1}{4}\alpha}}
-\frac{1}{16}(\alpha\gamma - \beta^2)
}}, \nonumber  \\  \\ 
&&\sigma' = \lambda 
\frac{\sqrt{\frac{1}{4}\alpha}\, 
\ch \tau\sqrt{\frac{1}{4}\alpha}}{\sh \tau\sqrt{\frac{1}{4}\alpha}}, \\[5pt]
&&\tau' = \lambda
\frac{\sqrt{\frac{1}{4}\gamma}\, 
\ch \sigma\sqrt{\frac{1}{4}\gamma}}{\sh \sigma\sqrt{\frac{1}{4}\gamma}},
\end{eqnarray}
where the symbols `$\ch$' and `$\sh$' are abbreviations of `$\cosh$' and `$\sinh$', respectively.
If we set $\sigma = \tau =1$ and slightly rewrite the expressions, we obtain 
(\ref{BCH_closed1})-(\ref{BCH_closed3}). 
If we further take $\alpha = \gamma = 0$ and set $\beta= 2a_{AB}P_{AB}$, we obtain 
 (\ref{XY_BCH1})-(\ref{XY_BCH2}).
Although the above derivation depends on a particular basis,
the result does not rely on any particular basis, therefore, it is applicable to 
any representations of $X$ and $Y$
which satisfy the algebra (\ref{XXY3_app})-(\ref{YYX3_app}).

\subsection{Application to group commutator}

\indent

Although not utilized in the main body of the paper,
it is worthwhile to note that the above-derived BCH closed forms 
can be applied recurcively.
As one of such examples, 
let us take $A$ and $B$ which satisfy
\begin{eqnarray}
[A, [A,B]] &=& \beta A, \ \ \ 
[B, [B,A]] \ =\  \beta B,
\end{eqnarray}
and try to calculate the  following type of group commutator, 
\begin{eqnarray}
e^{A}e^{B}e^{-A}e^{-B}, \label{G_Comm}
\end{eqnarray}
which is expected to be reduced, in the $\beta \rightarrow 0$ limit, as
\begin{eqnarray}
e^{A}e^{B}e^{-A}e^{-B} &=& e^{[A,B]}.
\end{eqnarray}
By applying $\alpha = \gamma = 0$ and $X=A$, $Y=B$
 in (\ref{XXY3}) and (\ref{YYX3}),
we obtain
\begin{eqnarray}
e^{A}e^{B}e^{-A}e^{-B} = e^{A'}  e^{B'},
\end{eqnarray}
where $A'$ and $B'$ are given by
\begin{eqnarray}
A' &=& F(0, \beta, 0) \Bigl( A+B+\frac{1}{2}[A,B] \Bigr), \ \ \
B' \ =\  F(0, \beta, 0) \Bigl( -A-B+\frac{1}{2}[A,B] \Bigr),
\end{eqnarray}
with $F(0, \beta, 0)$ given by setting $\alpha = \gamma =0$
in (\ref{BCH_closed2}), or explicitly,
\begin{eqnarray}
F(0, \beta, 0) &=& \frac{\log \Bigl(1-\frac{\beta}{4}+\sqrt{(1-\frac{\beta}{4})^2-1}\Bigr)}
{\sqrt{(1-\frac{\beta}{4})^2 -1}}.  
\end{eqnarray}
Noticing that commutation relations among $A'$ and $B'$ give 
\begin{eqnarray}
[A',B'] &=& F(0, \beta, 0)^{2}\beta\, (A-B),
\\[5pt]
[A',[A',B']] &=&  \gamma' B' + \beta' A', \\[5pt]
[B',[B',A']] &=& \gamma' A' + \beta' B', 
\end{eqnarray}
where $\beta'$ and $\gamma'$ are given by
in terms of  $\beta$ and $F(0, \beta, 0)$ as,
\begin{eqnarray}
\beta' &=& -F(0, \beta, 0)^{2}\beta\, (2+\frac{1}{4}\beta), \ \ \ 
\gamma' \ =\ -F(0, \beta, 0)^{2}\beta\, (2-\frac{1}{4}\beta),
\end{eqnarray}
we then obtain for the group commutator (\ref{G_Comm}), 
\begin{eqnarray}
e^{A}e^{B}e^{-A}e^{-B} &=& 
\exp \Biggl[ F(\gamma', \beta', \gamma') \Big(G(\gamma') A' 
+G(\gamma') B' +\frac{1}{2}[A',B']\Bigr) \Biggr]  \\[5pt]
&=& 
\exp \Biggl[ F(\gamma', \beta', \gamma') \Big(G(\gamma') F(0, \beta, 0)\,
[A,B] +\frac{1}{2}F(0, \beta, 0)^{2}\beta\, (A-B)\Bigr) \Biggr],\ \ \ 
\end{eqnarray}
where $F(\gamma', \beta', \gamma',)$ and $G(\gamma')$ are given by
setting $\alpha = \gamma = \gamma'$, $\beta = \beta'$
in (\ref{BCH_closed2}) and (\ref{BCH_closed3}), or exlicitly,
\begin{eqnarray}
F(\gamma', \beta', \gamma') &=& \frac{\log \Bigl( \cosh^2 \sqrt{\frac{1}{4}\gamma'} -\frac{\beta'}{\gamma'}\sinh^2\sqrt{\frac{1}{4}\gamma'} + \sqrt{\bigl(\cosh^2 \sqrt{\frac{1}{4}\gamma'} -\frac{\beta'}{\gamma'}\sinh^2\sqrt{\frac{1}{4}\gamma'}\bigr)^2 -1}\Bigr)}
{\sqrt{\frac{1}{8}\gamma' (\gamma' - \beta')
 \frac{\cosh^2\sqrt{\frac{1}{4}\gamma'}}{\sinh^2\sqrt{\frac{1}{4}\gamma'}} 
-  \frac{1}{16}(\gamma'^2 - \beta'^2)
}},\ \ \ \ \ \ \ \ \ \  \\[5pt]
G(\gamma') &=&  \frac{\sqrt{\frac{1}{4}\gamma'}\cosh\sqrt{\frac{1}{4}\gamma'}}{\sinh\sqrt{\frac{1}{4}\gamma'}}. 
\end{eqnarray}

\end{document}